%% file: static_dynamic_rpca_procIEEE_7.tex
\documentclass[10pt]{IEEEtran} 

\usepackage{etex}
\usepackage[utf8]{inputenc} 
\usepackage[T1]{fontenc}    
\usepackage{hyperref}       
\usepackage{url}            
\usepackage{amsfonts}       
\usepackage{microtype}      
\usepackage{hyperref}
\usepackage{pgfplots}
\usepackage{amsmath, amssymb, amsthm, algorithm, algpseudocode,epstopdf} 
\usepackage{mathtools}
\usepackage{bm}
\usepackage{epsfig,color,epstopdf,graphicx,multirow,array,bbm}
\usepackage{subcaption}

\usepackage{tabu,booktabs}

\renewcommand{\subsubsection}[1]{\noindent {\bf #1. }}

\input{tikz_def_proc} 

\input{Tex_Commands}

\renewcommand{\xmint}{s_{\min}}
\newcommand{\zz}{\epsilon}
\renewcommand{\dif}{\Delta}
\newcommand{\bpsi}{\bm\Psi}

\begin{document}

\title{Static and Dynamic Robust PCA and Matrix Completion: A Review}
\author{Namrata Vaswani and Praneeth Narayanamurthy
\\ Iowa State University, USA
}

\maketitle


\begin{abstract}
Principal Components Analysis (PCA) is one of the most widely used dimension reduction techniques. Robust PCA (RPCA) refers to the problem of PCA when the data may be corrupted by outliers. Recent work by Cand{\`e}s, Wright, Li, and Ma defined RPCA as a problem of decomposing a given data matrix into the sum of a low-rank matrix (true data) and a sparse matrix (outliers). The column space of the low-rank matrix then gives the PCA solution. This simple definition has lead to a large amount of interesting new work on provably correct, fast, and practical solutions to RPCA.
More recently, the dynamic (time-varying) version of the RPCA problem has been studied and a series of provably correct, fast, and memory efficient tracking solutions have been proposed. Dynamic RPCA (or robust subspace tracking) is the problem of tracking data lying in a (slowly) changing subspace, while being robust to sparse outliers.
This article provides an exhaustive review of the last decade of literature on RPCA and its dynamic counterpart (robust subspace tracking), along with describing their theoretical guarantees, discussing the pros and cons of various approaches, and providing empirical comparisons of performance and speed.

A brief overview of the (low-rank) matrix completion literature is also provided (the focus is on works not discussed in other recent reviews). This refers to the problem of completing a low-rank matrix when only a subset of its entries are observed. It can be interpreted as a simpler special case of RPCA in which the indices of the outlier corrupted entries are known. %
\end{abstract}



\section{Introduction} 

Principal Components Analysis (PCA) is one of the most widely used dimension reduction techniques. It is often the preprocessing step in a variety of scientific and data analytics' applications. Some modern examples include data classification, face recognition, video analytics, recommendation system design and understanding social network dynamics. 
PCA finds a small number of orthogonal basis vectors along which most of the variability of the dataset lies.
Given an $n \times d$ data matrix $\Y$, {\em $r$-PCA} finds an $n \times r$ matrix with orthonormal columns $\Phat_M$ that solves
$
\arg\min_{\tilde\P: \tilde\P'\tilde\P = \I} \|\Y - \tilde\P \tilde\P' \Y\|_2.
$
For dimension reduction, one projects $\Y$ onto $\Span(\Phat_M)$. 
By the Eckart-Young theorem \cite{eckart}, PCA is easily solved via singular value decomposition (SVD), i.e., $\Phat_M$ is  given by the left singular vectors of $\Y$ corresponding to the largest $r$ singular values (henceforth referred to as ``top $r$ singular vectors''). Here and below, $'$ denotes matrix transpose and $\I$ denotes the identity matrix.

The observed data matrix $\Y$ is usually a noisy version of an unknown true data matrix, which we will denote by $\L$. The real goal is usually to find the principal subspace of $\L$. $\L$ is assumed to be either exactly or approximately low-rank. Suppose it is exactly low-rank and let $\rmat$ denote its rank. 
If $\Y$ is a relatively clean version of $\L$, $\Phat_M$ is also a good approximation of the principal subspace of $\L$, denoted $\P$. However, if $\Y$ is a highly noisy version of $\L$ or is corrupted by even a few outliers, $\Phat_M$ is a bad approximation of $\P$. Here, ``good approximation" means, $\Span(\Phat_M)$ is close to $\Span(\P)$.
Since many modern datasets are acquired using a large number of inexpensive sensors, outliers are becoming even more common in modern datasets. They occur due to various reasons such as node or sensor failures, foreground occlusion of video sequences, or anomalies on certain nodes of a network.  This harder problem of PCA for outlier corrupted data is called {\em robust PCA} \cite{RP_Campell,RS_Huber,RPCAPP_Croux,FRPCA_Hubert,rpca_neu,Torre03aframework} 





In recent years, there have been multiple attempts to qualify the term ``outlier'' leading to various formulations for robust PCA (RPCA). The most popular among these is the idea of modeling outliers as additive sparse corruptions which was popularized in the work of Wright and Ma \cite{error_correction_PCP_l1,rpca_0}. This models the fact that outliers occur infrequently and only on a few indices of a data vector, but allows them to have any magnitude. Using this, the recent work of Cand{\`e}s, Wright, Li, and Ma \cite{rpca_0,rpca} defined RPCA as the problem of decomposing a given data matrix, $\Y$, into the sum of a low rank matrix, $\L$, and a sparse matrix (outliers' matrix), $\X$. The column space of $\L$ then gives the PCA solution.
While RPCA was formally defined this way first in \cite{rpca_0,rpca}, an earlier solution approach that implicitly used this idea was \cite{Torre03aframework}.

Often, for long data sequences, e.g., long surveillance videos, or long dynamic social network connectivity data sequences, if one tries to use a single lower dimensional subspace to represent the entire data sequence, the required subspace dimension may end up being quite large. This is problematic because (i) it means that PCA does not provide sufficient dimension reduction, (ii) the resulting data matrix may not be sufficiently low-rank, and this, in turn, reduces the outlier tolerance of static RPCA solutions, and (iii) it implies increased computational and memory complexity.
In this case, a better model is to assume that the data lies in a low-dimensional subspace that can change over time, albeit gradually. 
The problem of tracking data lying in a (slowly) changing subspace while being robust to additive sparse outliers is referred to as {\em ``robust subspace tracking''} or {\em ``dynamic RPCA''} \cite{rrpcp_allerton,rrpcp_tsp,rrpcp_perf,rrpcp_aistats,rrpcp_medrop}. In older work, it was also incorrectly called {\em ``recursive robust PCA'' or ``online robust PCA''} .

The current article provides a detailed review of the literature on both RPCA and dynamic RPCA (robust subspace tracking) that relies on the above sparse+low-rank matrix decomposition (S+LR) definition.  The emphasis is on simple and provably correct approaches. A brief overview of the low-rank matrix completion (MC) literature and of dynamic MC, or, equivalently subspace tracking (ST) with missing data is also provided. MC refers to the problem of completing a low-rank matrix when only a subset of its entries can be observed. We discuss it here because it can be interpreted as a simpler special case of RPCA in which the indices of the outlier corrupted entries are known.
A detailed review of MC and the more general problem of low rank matrix recovery from linear measurements is provided in \cite{davenport_review}. A detailed discussion of ST (including ST with missing data) is given in \cite{chi_review}. Another way to define the word ``outlier'' is to assume that either an entire data vector is an outlier or it is an inlier. In modern literature, this is referred to as ``robust subspace recovery'' \cite{novel_m_estimator}. This is reviewed in \cite{lerman_review}. A magazine-level overview of the entire field of RPCA including robust subspace recovery is provided in \cite{rrpcp_review}.%




A key motivating application for RPCA and robust subspace tracking (RST) is video layering (decompose a given video into a ``background'' video and a ``foreground'' video) \cite{Torre03aframework,rpca}. We show an example in Fig. \ref{fig_video}. While this is an easy problem for videos with nearly static backgrounds, the same is not true for dynamically changing backgrounds. For such videos, a good video layering solution can simplify many downstream computer vision and video analytics' tasks. For example, the foreground layer directly provides a video surveillance and an object tracking solution,  the background layer and its subspace estimate are directly useful in video background-editing or animation applications; video layering can also enable or improve low-bandwidth video chats (transmit only the layer of interest), layer-wise denoising \cite{rrpcp_denoise,rrpcp_denoise_jp} or foreground activity recognition. RPCA is a good video layering solution when the background changes are gradual (typically valid for static camera videos) and dense (not sparse), while the foreground consists of  one or more moving persons/objects that are not too large. An example is background variations due to lights being turned on and off shown in Fig. \ref{realvid}. With this, the background video (with each image arranged as a column) is well modeled as the dense low rank matrix while the foreground video is well modeled as a sparse matrix with high enough rank. Thus the background corresponds to $\L$, while the difference between foreground and background videos on the foreground support forms the sparse outliers $\S$. Other applications include region of interest  detection and tracking from full-sampled or under-sampled dynamic MRI sequences \cite{candes_mri}; detection of anomalous behavior in dynamic social networks \cite{selin_reprocs}, or in computer networks \cite{mateos_anomaly}; recommendation system design and survey data analysis when there are outliers due to lazy users and typographical errors \cite{rpca}.

A motivating video analytics application for matrix completion is the above problem when the foreground occlusions are easily detectable (e.g., by simple thresholding). Related problems include  dense low-rank image or video recovery when some pixels are missing, e.g., due to data transmission errors/erasures (erasures are transmission errors that are reliably detectable so that the missing pixel indices are known); and image/video inpainting when the underlying image or video (with images vectorized as its columns) is well modeled as being dense and low-rank. Another motivating application is recommendation system design (without lazy users) \cite{matcomp_candes}. This assumes that user preferences for a class of items, say movies, are governed by a much smaller number of factors than either the total number of users or the total number of movies. The movie ratings' matrix is incomplete since a given user does not rate all movies.

\input{proc_intro_figs}

\subsection{Article organization}
This article is organized as follows. We give the problem setting for both RPCA and dynamic RPCA next, followed by discussing identifiability and other assumptions, and then give the matric completion problem formulation. Section II describes the original (static) RPCA solutions. Section III discusses two key approaches to dynamic robust PCA / robust subspace tracking. The algorithm idea, why it works, and its theoretical guarantees (where they exist) are summarized.
Section IV discusses the pros and cons of all the approaches that are described in detail in this article.
In Section V, we provide a brief summary of solutions for matrix completion and for subspace tracking with missing data. 
Section VI provides empirical performance and speed comparisons for original and dynamic RPCA. We conclude in Section VII with a detailed discussion of open questions for future work.

\subsection{Notation}
We use $'$ to denote matrix transpose; $\I$ denotes the identity matrix; we use $[a, b]$ to refer to all integers between $a$ and $b$, inclusive, $[a,b): = [a,b-1]$. The notation $\M_\T$ denotes a sub-matrix of $\M$ formed by its columns indexed by entries in the set $\T$. Also, $\|.\|$ denotes the induced $l_2$ norm.

In this article $\P$ or $\Phat$ with different subscripts always refer to ``basis matrices'' (tall matrices with orthonormal columns). These are used to denote the subspaces spanned by their columns. When we say $\Phat_M$ is a good approximation of $\P_L$, we mean that the corresponding subspaces are close. We use $\SE(\Phat,\P):=\|(\I - \Phat \Phat')\P\|$ to measure the Subspace Error (distance). This measures the sine of the maximum principal angle between the subspaces. It is symmetric when the two subspaces have equal dimension.

\subsection{RPCA}
Given an $n \times \tmax$ observed data matrix, 
\[
\M := \L + \S + \W,
\]
where $\L$ is a low rank matrix (true data), $\S$ is a sparse matrix (outlier matrix), and $\W$ is small unstructured noise/modeling error, the goal is to estimate $\L$, and hence its column space.
We use $\rmat$ to denote the rank of $\L$. The maximum fraction of nonzeros in any row (column) of the outlier matrix $\S$ is denoted by $\outfracrow$ ($\outfraccol$).

\subsection{Robust Subspace Tracking (RST) or dynamic RPCA}
At each time $t$, we get a data vector $\yt \in \Re^n$ that satisfies%
\bea
\mt := \lt + \s_t + \v_t, \text{ for } t = 1, 2, \dots, \tmax. \nn
\label{orpca_eq}
\eea
where  $\v_t$ is small unstructured noise, $\xt$ is the sparse outlier vector, and $\lt$ is the true data vector that lies in a fixed or slowly changing low-dimensional subspace of $\Re^n$, i.e., $\lt = \P_{(t)} \a_t$ where $\P_{(t)}$ is an $n \times r$ {\em basis matrix}\footnote{matrix with mutually orthonormal columns} with $r \ll n$ and with $\|(\I - \P_{(t-1)}\P_{(t-1)}{}')\P_{(t)}\|$ small compared to $\|\P_{(t)}\|=1$.
We use $\T_t$ to denote the support set of $\xt$.
%
%
Given an initial subspace estimate, $\Phat_{0}$, the goal is to track $\Span(\P_{(t)})$ and $\lt$ either immediately or within a short delay  \cite{rrpcp_perf,rrpcp_dynrpca,rrpcp_medrop}. A by-product is that $\lt$, $\x_t$, and $\T_t$ can also be tracked on-the-fly.

The initial subspace estimate, $\Phat_0$, can be computed by using only a few iterations of any of the static RPCA solutions (described below), e.g., PCP \cite{rpca} or AltProj \cite{robpca_nonconvex}, applied to the first $t_\train$ data samples $\Y_{[1,t_\train]}$. Typically,  $t_\train=Cr$ suffices.
Alternatively, in some applications, e.g., video surveillance, it is valid to assume that outlier-free data is available. In these situations, simple SVD can be used too.

Technically, {\em dynamic RPCA} refers to the offline version of the RST problem. Define matrices $\L,\S,\V,\Y$ with $\L = [\l_1,\l_2, \dots \l_{\tmax}]$ and $\Y,\S,\V$ similarly defined.  The goal is to recover the matrix $\L$ and its column space with $\epsilon$ error.

\subsection{Identifiability and other assumptions}
The above problem definitions do not ensure identifiability since either of $\L$ or $\S$ can be both low-rank and sparse. 
One way to ensure that $\L$ is not sparse is by requiring that its left and right singular vectors are dense or ``incoherent'' w.r.t. a sparse vector \cite{rpca}. We define this below.
\begin{definition}[$\mu$-Incoherence/Denseness]
We say that an $n \times r$ basis matrix  (matrix with mutually orthonormal columns) $\P$  is $\mu$-incoherent if
\bea
\max_{i=1,2,\dots, n} \|\P^{(i)}\|^2 \le {\mu \frac{\rmat}{n}} \nn
\eea
where $\mu \ge 1$ is  called the (in)coherence parameter that quantifies the non-denseness  of $\P$.
Here $\P^{(i)}$ denotes the $i$-th row of $\P$.
\end{definition}

We can ensure that $\S$ is not low-rank in one of two ways. The first is to impose upper bounds on $\outfracrow$ and $\outfraccol$. The second is to assume that the support set of $\S$ is generated uniformly at random (or according to the independent identically distributed (iid) Bernoulli model) and then to just bound the total number of its nonzero entries. The uniform random or iid Bernoulli models ensure roughly equal nonzero entries in each row/column. 

Consider the Robust Subspace Tracking problem. The most general nonstationary model that allows the subspace to change at each time is not even identifiable since at least $r$ data points are needed to compute an $r$-dimensional subspace even in the noise-free full data setting. 
%
%
One way \cite{rrpcp_perf,rrpcp_aistats,rrpcp_medrop} to ensure identifiability of the changing subspaces is to assume that they are piecewise constant, i.e., that
\[
\P_{(t)} = \P_{j} \text{ for all } t \in [t_j, t_{j+1}),  \  j=1,2,\dots, J,
\]
with $t_0=1$ and $t_{J+1}=\tmax$. 
With the above model, in general, $\rmat = r J$ (except if subspace directions are repeated more than once, or if only a few subspace directions change at some change times).

\subsection{Matrix Completion}
(Low Rank) Matrix Completion (MC) refers to the problem of completing a rank $r$ matrix $\L$ from a subset of its entries. We use $\Omega$ to refer to the set of indices of the observed entries of $\L$ and we use the notation $\pP_\Omega(\M)$ to refer to the matrix formed by setting the unobserved entries to zero. Thus, given
\[
\Y: = \pP_\Omega(\L)
\]
the goal of MC is to recover $\L$ from $\Y$. {\em The set $\Omega$ is known.} To interpret this as a special case of RPCA, notice that one can rewrite $\Y$ as $\Y = \L - \pP_{\Omega^c}(\L)$ where $\Omega^c$ refers to the complement of the set $\Omega$. By letting $\S = - \pP_{\Omega^c}(\L)$, this becomes a special case of RPCA.

{\em Identifiability. } Like RPCA, this problem is also not identifiable in general. For example, if $\L$ is low-rank and sparse and if one of its nonzero entries is missing there is no way to ``interpolate'' the missing entry from the observed entries without extra assumptions. This issue can be resolved by assuming that the left and right singular vectors of $\L$ are $\mu$-incoherent as defined above. In fact incoherence was first introduced for the MC problem in \cite{matcomp_candes}, and later used for RPCA.
Similarly, it is also problematic if the set $\Omega$ contains all entries corresponding to just one or two columns (or rows) of $\L$; then, even with the incoherence assumption, it is not possible to correctly ``interpolate'' all the columns (rows) of $\L$. This problem can be resolved by assuming that $\Omega$  is generated uniformly at random (or according to the iid Bernoulli model) with a lower bound on its size.
For a detailed discussion of this issue, see \cite{matcomp_candes,matcomp_noise}.

{\em``Robust MC'' (RMC) or ``Robust PCA with Missing Data''} \cite{rmc,rmc_gd} is an extension of both RPCA and MC. It involves recovering $\L$ from $\Y$  when $\Y = \pP_\Omega(\L + \S)$. Thus the entries are corrupted and not all of them are even observed. In this case there is no way to recover $\S$ of course. Also, the only problematic outliers are the ones that correspond to the observed entries since $\Y = \pP_\Omega(\L) + \pP_\Omega(\S)$.

{\em Dynamic MC} is the same as the problem of {\em subspace tracking with missing data (ST-missing)}. This can be defined in a fashion analogous to the RST problem described above. Similarly for dynamic RMC.

\subsection{Other Extensions}
In many of the applications of RPCA, the practical goal is often to find the outlier or the outlier locations (outlier support). For example, this is often the case in the video analytics application. This is also the case in the anomaly detection application. In these situations, robust PCA should really be called {\em ``robust sparse recovery'', or ``sparse recovery in large but structured noise''}, with ``structure'' meaning that the noise lie in a fixed or slowly changing low-dimensional subspace \cite{rrpcp_perf}. 
%
%
Another useful extension is {\em undersampled or compressive RPCA or robust Compressive Sensing (CS)} \cite{rrpcp_allerton11,SpaRCS,compressivePCP,rpca_Giannakis,candes_mri}. 
Instead of observing the matrix $\M$, one only has access to a set of $m < n $ random linear projections of each column of $\M$, i.e., to $\bm{Z} = \bm{A} \M$ where $\bm{A}$ is a fat matrix. An important application of this setting is in dynamic MRI imaging when the image sequence is modeled as sparse + low-rank \cite{candes_mri}. An alternative formulation is Robust CS where one observes $\bm{Z}:=\bm{A} \X + \L$ \cite{rrpcp_allerton11,rrpcp_tsp,SpaRCS,rpca_Giannakis} and the goal is to recover $\X$ while being robust to $\L$. This would be dynamic MRI problem if the low rank corruption $\L$ is due to measurement noise.

\section{RPCA solutions}
 Before we begin, we should mention that the code for all the methods described in this section is downloadable from the github library of Andrew Sobral \cite{lrslibrary2015}. The link is \url{https://github.com/andrewssobral/lrslibrary}.

Also, in the guarantees given in this article, for simplicity, the condition number is assumed to be constant, i.e., $O(1)$, with $n$.

\subsection{Principal Component Pursuit (PCP): a convex programming solution} \label{pcp_section}
The first provably correct solution to robust PCA via S+LR was introduced in parallel works by Cand{\`e}s, Wright, Li, and Ma \cite{rpca} (where they called it a solution to robust PCA) and by Chandrasekharan et al. \cite{rpca2}. Both proposed to solve the following convex program which was referred to as Principal Component Pursuit (PCP) in \cite{rpca}:
\[
\min_{\tL, \tS}  \|\tL\|_* + \lambda \|\tS\|_{vec(1)}  \text{ subject to } \M = \tL + \tS
\]
Here $\|\A\|_{vec(1)}$ denotes the vector $l_1$ norm of the matrix $\A$ (sum of absolute values of all its entries) and $\|\A\|_*$ denotes the nuclear norm  (sum of its singular values).
PCP is the first known polynomial time solution to RPCA that is also provably correct. The two parallel papers \cite{rpca,rpca2} used different approaches to arrive at a correctness result for it. The result of \cite{rpca_zhang} improved that of \cite{rpca2}.


Suppose that PCP can be solved exactly. Denote its solutions by $\Lhat, \Shat$. The result of \cite{rpca} says the following.
\begin{theorem}
Let $\L \svdeq \LU \bm\Sigma \LV'$ be its reduced SVD.
If $\V=0$,
\ben
\item $\LU$ is $\mu$-incoherent, $\LV$ is $\mu$-incoherent,
\item $\LU$ and $\LV$ are $\mu$-strong-incoherent, i.e. satisfy 
\bea
\max_{i=1,2,\dots, n, j=1,2,\dots,\tmax} |(\LU\LV')_{i,j}| \le \sqrt{\mu \frac{\rmat}{n \tmax}} \ \ \nn 
\label{strong_incoh}
\eea
\item support of $\S$ is generated uniformly at random,
\item the support size of $\S$, denoted $m$,  and the  rank of $\L$, $\rmat$, satisfy:
$\frac{m}{n \tmax} \le c $ and $\rmat \le \frac{c \min(n,\tmax)}{\mu (\log n )^2 }$, 
\een
the parameter $\lambda = 1/\sqrt{\max(n,\tmax)}$ \footnote{Notice that this requires no knowledge of model parameters.},
then, with probability at least $1- c n^{-10}$, the PCP convex program with $\lambda = 1/\sqrt{\min(n,d)}$ returns $\Lhat=\L$ and $\Shat=\S$.%
%
\label{thm_candes}
\end{theorem}
The second condition (strong incoherence) requires that the inner product between a row of $\LU$ and a row of $\LV$ be upper bounded. Observe that the required bound is $1/\sqrt{\rmat}$ times what left and right incoherence would imply (by using Cauchy-Schwartz inequality). This is why it is a stronger requirement.

The guarantee of \cite{rpca_zhang}, which improved the result of \cite{rpca2}, says the following. We give below a simpler special case of \cite[Theorem 2]{rpca_zhang} applied with $\rho = \sqrt{n/d}$ and for the exact (noise-free) PCP program given above\footnote{The parameter $\rho$ is defined in the first column of page 7223 of \cite{rpca_zhang} as the balancing parameter to accommodate disparity between the number of rows and columns of the data matrix. In our notation, the data matrix $\Y$ is of size $n \times d$, thus, setting $\rho = \sqrt{n/d}$ ensures that $\alpha(\rho)$ of \cite{rpca_zhang} is proportional to $\max(\outfracrow,\outfraccol)$.}.
\begin{theorem}
Let $\L \svdeq \LU \bm\Sigma \LV'$. If $\V=0$,
\ben
\item each entry of $\LU$ is of order $\sqrt{C/n}$ and each entry of $\LV$ is of order $\sqrt{C/d}$,
\item $\max(\outfracrow,\outfraccol) \le \frac{c}{ \mu \rmat }$,
\een
the parameter $\lambda$ lies in a certain range\footnote{The bounds on $\lambda$ depend on $\max(\outfracrow,\outfraccol)$.},
then $\Lhat=\L$ and $\Shat=\S$.
\label{thm_zhang}
\end{theorem}

Theorem \ref{thm_zhang} does not assume a model on outlier support, but because of that, it needs a much tighter bound of $O(1/\rmat)$ on outlier fractions. Theorem \ref{thm_candes} assumes uniform random outlier support, along with the support size $m$ bounded by $c n \tmax$. For large $n,\tmax$, this is approximately equivalent to allowing $\max(\outfracrow,\outfraccol) \le c$.
This is true because for large $n,\tmax$, with high probability (w.h.p.), (i) uniform random support with size $m$ is nearly equivalent\footnote{recovery under one model implies recovery under the other model with same order of probability} to Bernoulli support with probability of an index being part of the support being $\rho = m/(n \tmax)$ \cite[Appendix 7.1]{rpca}; and (ii) with the Bernoulli model, $\max(\outfracrow,\outfraccol)$ is close to $\rho$ (follows using Hoeffding inequality for example).

{\em Why PCP works. } It is well known from compressive sensing literature (and earlier) that the vector $l_1$ norm serves as a convex surrogate for the support size of a vector, or of a vectorized matrix \cite{candes_rip}. In a similar fashion, the nuclear norm serves as a convex surrogate for the rank of a matrix \cite{matcomp_first,matcomp_candes}. Thus, while the program that tries to minimize the rank of $\tL$ and sparsity of $\tS$ involves an impractical combinatorial search, PCP is convex and solvable in polynomial time \cite{rpca}.


\subsection{Alternating Minimization (AltProj): a non-convex solution}



Convex optimization programs as solutions to various originally non-convex problems (e.g., robust PCA, sparse recovery, low rank matrix completion, phase retrieval) are, by now, well understood. They are easy to come up with (often), solvable in polynomial time (polynomial in the data size), and allow one to come up with strong guarantees with minimal sample complexity. While polynomial complexity is better than exponential, it is often too slow for today's big datasets. Moreover, the number of iterations needed for a convex program solver to get to within an $\epsilon$ ball of the true solution of the convex program is $O(1/\epsilon)$ and thus the typical complexity for a PCP solver is $O(n d^2/\epsilon)$ \cite{robpca_nonconvex}.
To address this limitation, in more recent works \cite{lowrank_altmin,robpca_nonconvex,pr_altmin,twf,rwf,lrpr_tsp}, authors have developed provably correct alternating minimization (alt-min) or projected gradient descent (GD) solutions that are provably much faster, but still allow for the same type of performance guarantees.
Both alt-min and GD have been used for a long time as practical heuristics for trying to solve various non-convex programs. The initialization either came from other prior information, or multiple random initializations were used to run the algorithm and the ``best'' final output was picked.
%
The new ingredient in these provably correct alt-min or GD solutions is a carefully designed initialization scheme that already outputs an estimate  that is ``close enough'' to the true one.
Since these approaches do not use convex programs, they have been labeled as ``non-convex'' solutions.

For RPCA, the first such provably correct solution was Alternating-Projection (AltProj) \cite{robpca_nonconvex}. The idea itself is related to that of an earlier algorithm called GoDec \cite{godec}. In fact the recursive projected compressive sensing (ReProCS) \cite{rrpcp_allerton,rrpcp_allerton11,rrpcp_perf} approach is an even earlier approach that also used a similar idea.
AltProj is summarized in Algorithm \ref{altproj_algo}. It alternates between estimating $\L$ with $\S$ fixed at its previous estimate, followed by projection onto the space of low-rank matrices, and then a similar procedure for $\S$.
Theorem 2 of \cite{robpca_nonconvex} says the following.
\begin{theorem}
Let $\L \svdeq \LU \bm\Sigma \LV'$. If $\V=0$,
\ben
\item $\LU$, $\LV$ are $\mu$-incoherent,
\item $\max(\outfracrow,\outfraccol) \le \frac{c}{ \mu \rmat }$,
\een
algorithm parameters are appropriately set\footnote{Need knowledge of $\mu$ and $\rmat$.}, then AltProj returns $\Lhat, \Shat$ that satisfy $\|\Lhat-\L\|_F \le \epsilon$, $\|\Shat-\S\|_{\max} \le \epsilon$, and $\supp(\Shat) \subseteq \supp(\S)$. Here $\max$ refers to the maximum nonzero entry of the matrix.

If $\V \neq 0$, but $\|\V\|_F^2 \le C \epsilon^2$, all guarantees remain the same.

AltProj needs time of order $O( n d \rmat^2 \log(1/\epsilon))$ and memory of $O(nd)$ to achieve above error.
\end{theorem}
Notice that even in the $\V=0$ case the above result only guarantees recovery with $\epsilon$ error while PCP seems to guarantee ``exact'' recovery.
This guarantee may seem weaker than that for PCP, however it actually is not. The reason is that any solver (the iterative algorithm for finding a solution) of the convex program PCP is only guaranteed to get you within $\epsilon$ error of the true solution of PCP in a finite number of iterations.


{\em Why AltProj works. }
To understand why AltProj works, consider the rank one case. As also explained in \cite{rrpcp_review} and in the original paper \cite{robpca_nonconvex}, once the largest outliers are removed, it is expected that projecting onto the space of rank one matrices returns a reasonable rank one approximation of $\L$, $\Lhat_1$. This means that the residual $\M - \Lhat_1$ is a better estimate of $\S$ than $\M$ is. Because of this, it can be shown that $\Shat_1$ is a better estimate of $\S$ than $\Shat_0$  and so the residual $\M - \Shat_1$ is a better estimate of $\L$ than $\M - \Shat_0$. This, in turn, means $\Lhat_2$ will be a better estimate of $\L$ than $\Lhat_1$ is.
The proof that the initial estimate of $\L$ is good enough  relies on incoherence of left and right singular vectors of $\L$ and the fact that no row or column has too many outliers.
These two facts are also needed to show that each new estimate is better than the previous.

\begin{algorithm}[t!]
\caption{\small{AltProj algorithm}}
AltProj for rank-1 matrix $\L$ ($HT$ denotes the hard thresholding operator, see discussion in the text):
\bi
\item Initialize
\bi
\item set $\Lhat^0 = 0$;
\item threshold out large entries from $\M$ to get $\Shat^0 = HT_{\beta \sigma_1}(M)$, where $\sigma_1 = \sigma_1(\M)$
\ei
\item
For Iterations $t=1$ to $T: = c \log(1/\epsilon)$ do
\bi
\item $\Lhat^t = \mathcal{P}_1 (\M - \Shat^t)$ : project $\M - \Shat^t$ onto space of rank-1 matrices
\item $\Shat^t = HT_{\beta (\sigma_t + 0.5^t \sigma_{t-1})}(\M - \Lhat^t)$, $\sigma_t := \sigma_t(\M - \Shat^t)$
\ei
End For
\ei
For general rank $r$ matrices $\L$:
\bi
\item The algorithm proceeds in $r$ stages and does $T$ iterations for each stage.
\item Stage 1 is the same as the one above. In stage k, $\pP_1$ is replaced by $\pP_k$ (project onto space of rank $k$ matrices). The hard thresholding step uses a threshold of $\beta(\sigma_{k+1} + 0.5^t \sigma_k)$.
\ei
\label{altproj_algo}
\end{algorithm}

\subsection{Memory-Efficient Robust PCA (MERoP) via Recursive Projected Compressive Sensing: a non-convex and online solution} \label{merop}
ReProCS \cite{rrpcp_merop} is an even faster solution than AltProj, that is also online (after a batch initialization applied to the first $C r $ frames) and memory-efficient. In fact, it has near-optimal memory complexity of $O(n \rmat \log n \log(1/\epsilon))$. Its time complexity of just $O(n d \rmat \log(1/\epsilon))$ is the same as that of vanilla $r$-SVD (simple PCA).
Moreover, after initialization, it also has the best outlier tolerance: it tolerates $\outfracrow^\alpha \in O(1)$. But the tradeoff is that it needs to assume that (i) all $\lt$'s lie in either a fixed subspace or a slowly changing one, and that (ii) (most) outlier magnitudes are lower bounded. As we explain later both are natural assumptions for static camera videos.
It relies on the recursive projected compressive sensing (ReProCS) approach \cite{rrpcp_allerton11,rrpcp_perf} introduced originally to solve the dynamic RPCA problem. But, equivalently, it also solves the original RPCA problem with the extra assumption that the true data subspace is either fixed or changes slowly. The simplest ReProCS-based algorithm is explained and summarized later as Algorithm \ref{norst_basic} (ReProCS-NORST) given later in Sec. \ref{reprocs_details}. ReProCS starts with a rough estimate of the initial subspace ($\Span(\P_0)$), obtainable using a batch technique applied to an initial short sequence. At each time $t$, it iterates between a robust regression step (that uses columns of $\Phat_{(t-1)}$ as the regressors) to estimate $\xt$ and $\lt$, and a subspace update step (updates the subspace estimate every $\alpha$ frames via solving a PCA problem with using the last $\alpha$ $\lhat_t$'s as input data).




\subsection{Projected Gradient Descent (RPCA-GD): another non-convex batch solution}
An equally fast, but batch, solution that relied on projected gradient descent (GD), called RPCA-GD, was proposed in recent work \cite{rpca_gd}. If condition numbers are treated as constants, its complexity is $O(n d \rmat \log(1/\epsilon))$. This is the same as that of the ReProCS solution given above and hence also the same as that of vanilla $r$-SVD for simple PCA. To achieve this complexity, like ReProCS, RPCA-GD also needs an extra assumption: it needs a $\sqrt{r}$ times tighter bound on outlier fractions than what AltProj needs. 

Projected GD is a natural heuristic for using GD to solve constrained optimization problems. To solve $\min_{x \in \cal{C}} f(x)$, after each GD step, projected GD projects the  output onto the set $\cal{C}$ before moving on to the next iteration. RPCA-GD rewrites $\L$ as $\L = \tU \tV'$ where  where $\tU$, $\tV$ are $n \times r$ and $\tmax \times r$ matrices respectively.
At each iteration, RPCA-GD involves one projected GD step for $\tU$, $\tV$, and $\S$ respectively. For $\tU$, $\tV$ the ``projection'' is onto the space of $\mu$-incoherent matrices, while for $\S$ it is onto the space of sparse matrices.
Corollary 1 of \cite{rpca_gd} guarantees the following.
\begin{theorem}
Consider RPCA-GD.
Let $\L \svdeq \LU \bm\Sigma \LV'$. If $\V=0$,
\ben
\item $\LU$ is $\mu$-incoherent, $\LV$ is $\mu$-incoherent,
\item $\max(\outfracrow,\outfraccol) \le {c}/{\mu \rmat^{1.5} }$,
\een
algorithm parameters are appropriately set\footnote{Need knowledge of $\rmat$, $\mu$, $\sigma_{\max}(\L)$, $\sigma_{\min}(\L)$},
then $\|\Lhat-\L\|_F \le \epsilon \sigma_{\max}(\L)$.

RPCA-GD needs time of order $O(n \tmax \rmat  \log(1/\epsilon))$and memory of $O(nd)$ to achieve the above error.
\end{theorem}

\subsection{Projected GD for RMC (NO-RMC): non-convex batch solution}
The authors of \cite{rmc_gd} develop a different type of projected GD solution for robust matrix completion (RMC), and  argue that, even in the absence of missing entries, the same algorithm provides the fastest solution to RPCA as long as the following extra assumption holds: the {\em data matrix is nearly square}. i.e. as long as $\tmax \approx n$. This can be a stringent requirement, e.g, for video data the number of frames $\tmax$ is often much smaller than image size $n$. 
\begin{theorem}
Consider NO-RMC.
Let $\L \svdeq \LU \bm\Sigma \LV'$. If $\V=0$,
\ben
\item $\tmax \le n$ and $n = O(\tmax)$,
\item $\LU$ is $\mu$-incoherent, $\LV$ is $\mu$-incoherent,
\item $\max(\outfracrow,\outfraccol) \le \frac{c}{\mu \rmat }$,
\item the observed entries' set $\Omega$ is generated according to the Bernoulli model with probability $p$ that satisfies
\[
p \ge C n \tmax \mu^4 n \rmat^2 \log^2 n  (\log (\mu^2 \rmat \sigma_{\max}(\L)/\epsilon) )^2,
\]
\een
algorithm parameters are appropriately set\footnote{Need knowledge of $\rmat$, $\mu$, $\sigma_{\max}(\L)$.},
then $\|\Lhat-\L\|_F \le \epsilon$.

NO-RMC needs time of order $O(n \rmat^2 \log^2 n \log(1/\epsilon))$ and memory of $O(nd)$ to achieve the above error.
\end{theorem}
The corollary for RPCA is an obvious one. To get a fast RPCA solution, NO-RMC deliberately undersamples the full matrix $\M$. Thus as long as $m,\tmax$ are of the same order, NO-RMC needs only $O(nr^2)$ time up to log factors making it the fastest solution for data matrices that are nearly square.


\section{Robust Subspace Tracking and Dynamic RPCA}
 Before we begin, we should mention that the code for all the methods described in this section is downloadable from the github library of Andrew Sobral \cite{lrslibrary2015}. The link is \url{https://github.com/andrewssobral/lrslibrary}.
%

Consider the robust subspace tracking (RST) or dynamic RPCA problem defined earlier. This assumes that the true data vectors $\lt$ are generated from a (slow) time-varying subspace rather than a fixed one.
As explained in the introduction, this model is most useful for long data sequences, e.g., long surveillance videos. The reason is if one tries to use a single lower dimensional subspace to represent the entire data sequence, the required subspace dimension may end up being quite large. 
One way to address this using RPCA solutions would be to split the data matrix into smaller pieces and apply PCP or any of the other static RPCA solutions on each piece individually. However this would mean that one is not jointly using all the available information. This results in significant loss in performance in settings where the data subspace changes slowly over time, so that the previously recovered piece of the data matrix contains a lot of information for correctly recovering the current piece.

In the subsections below, we describe two different approaches that address this issue.
The first is a mini-batch approach called modified-PCP. This modifies the PCP convex program to apply it to mini-batches of data, and each time also use knowledge of the previous subspace and slow subspace change. An alternate solution is called ReProCS that uses the previous subspace estimate and slow subspace change to enable recovery of $\xt$ and $\lt$ at each time $t$ via solving a robust regression problem followed by a subspace update step. This is online for recovery of $\xt$ and $\lt$ while being mini-batch for recovering the subspace estimates. Before we begin, we provide a quick discussion of modeling time-varying subspaces.

\subsection{Modeling time-varying subspaces}
 Even though the RPCA problem has been extensively studied in the last decade (as discussed above), there has been only a small amount of work on provable dynamic RPCA and robust subspace tracking (RST) \cite{rrpcp_perf,rrpcp_aistats,rrpcp_medrop}. The subspace tracking (ST) problem (without outliers), and with or without missing data, has been studied for much longer in both the control theory and the adaptive signal processing literature \cite{past,past_conv,adaptivesigproc_book,grouse,petrels,local_conv_grouse}. However, all existing guarantees are asymptotic results for the statistically stationary setting of data being generated from a {\em single unknown} subspace. Moreover, most of these also make assumptions on intermediate algorithm estimates, see Sec. \ref{subtrack}. 
Of course, as explained earlier, the most general nonstationary model that allows the subspace to change at each time is not even identifiable since at least $r$ data points are needed to compute an $r$-dimensional subspace even in the noise-free full data setting. 
%
%
In recent work \cite{rrpcp_perf,rrpcp_aistats,rrpcp_medrop,zhan_pcp_jp}, the authors have made the tracking problem identifiable by assuming a piecewise constant model on subspace change. With this, they are able to show in \cite{rrpcp_medrop} that it is possible to track the changing subspace to within $\epsilon$ accuracy as long as the subspace remains constant for at least $C r \log n \log (1/ \epsilon)$ frames at a time, and some other assumptions hold. Here and elsewhere the letters $c$ and $C$ is reused to denote different numerical constants in each use. We describe this work below.

 \subsection{Modified-PCP: Robust PCA with partial subspace knowledge}
A simple extension of PCP, called modified-PCP, provides a nice solution to the problem of RPCA with partial subspace knowledge \cite{zhan_pcp_jp}.
In many applications, e.g., face recognition, some training data for face images taken in controlled environments (with eyeglasses removed and no shadows) is typically  available. This allows one to get ``an'' estimate of the faces' subspace that serves as the partial subspace knowledge. 
%
To understand the modified-PCP idea, let $\G$ denote the basis matrix for this partial subspace knowledge.  If $\G$ is such that the matrix $(\I-\G\G')\L$ has rank significantly smaller than $\rmat$, then the following is a better idea than PCP:
\bea
 \text{min}_{\tL,\tS}    \|(\I - \G \G') \tL\|_* + \lambda \|\tS\|_1      \text{ s.t. }  \tL + \tS =  \M.
\label{modpcp}
\eea
The above solution was called {\em modified-PCP} because it was inspired by a similar idea, called modified-CS \cite{modcsjournal}, that solves the problem of compressive sensing (or sparse recovery) when partial support knowledge is available.
More generally, even if only the {\em approximate} rank of $(\I-\G\G')\L$ is much smaller, i.e., suppose that $\L = \G \A  + \L_\new + \W$ where $\L_\new$ has rank $r_\new \ll \rmat$ and $\|\W\|_F \le \zz$ is small, the following simple change to the same idea works:
\[
 \text{min}_{\tL_\new,\tS, \tilde{\A}}   \|\tL_\new\|_* + \lambda \|\tS\|_1    \text{ s.t. }  \| \M - \tS - \G \tilde\A - \tilde\L_\new \|_F \le \zz
\]
and output $\Lhat = \G \hat\A + \Lhat_\new$. We should mention that the same type of idea can also be used to obtain a Modified-AltProj or a Modified-NO-RMC algorithm. As with PCP, these will be significantly faster than the above modified-PCP convex program.

When solving the dynamic RPCA problem using modified-PCP, the subspace estimate from the previous set of $\alpha$ frames serves as the partial subspace knowledge for the current set. It is initialized using PCP. It has the following guarantee \cite{zhan_pcp_jp}.
 \begin{theorem}
Consider the dynamic RPCA problem with $\vt=0$. Split the data matrix into pieces $\L_1, \L_2, \dots, \L_J$ with $\L_j:=\L_{[t_j,t_{j+1})}$.  Recover $\L_j$ using the column space estimate of $\L_{j-1}$ as the partial subspace knowledge $\G$ and solving \eqref{modpcp}. With probability at least $1- c n^{-10}$, $\Lhat = \L$ if $t_j$'s are known (or detected using the ReProCS idea), and the following hold:
\ben
\item $\L_0$ is correctly recovered using PCP 

\item The subspace changes as $\P_j = [\P_{j-1}, \P_{j,\new}]$, where $\P_{j,\new}$ has $r_\new$ columns that are orthogonal to $\P_{j-1}$, followed by removing some directions.

\item Left incoherence holds for $\P_j$'s; right incoherence holds for $\LV_{j,\new}$; and strong incoherence holds for the pair $\P_{j,\new}, \LV_{j,\new}$. 
    If $\L_j \svdeq \P_j \bm\Sigma_j \LV_j'$, then $\LV_{j,\new}$ is the matrix of last $r_\new$ columns of $\LV_j$.

\item Support of $\S$ is generated uniformly at random.

\item The support size of the support of $\S$, $m$, and the rank of $\L$, $\rmat$, satisfy:
$\frac{m}{n \tmax} \le c $ and $\rmat \le \frac{c \min(n, (t_{j+1}-t_j))}{\mu (\log n )^2 }$. 
\een
\end{theorem}
The  Modified-PCP guarantee was proved using ideas borrowed from \cite{rpca} for PCP (PCP(C)). Hence it inherits most of the pros and cons of the PCP(C) guarantee. Thus, it also needs uniformly randomly generated support sets which is an unrealistic requirement. Moreover, its slow subspace change assumption is unrealistic. Third, it does not detect the subspace change automatically, i.e., it assumes $t_j$ is known.
All these limitations are removed by the ReProCS solution described next, however, it needs a lower bound on outlier magnitudes.

\begin{algorithm}[t!]
\caption{\small{{\bf ReProCS-NORST} simplified: Only for simplicity of explanation, this assumes $t_j$'s known. Both guarantees and experiments use the automatic version given in \cite[Algorithm 2]{rrpcp_medrop}.
\\ Notation: $\hat{\L}_{t; \alpha} := [\lhat_{t-\alpha + 1}, \cdots, \lhatt]$ and $\SVD_r[\M]$ refers to the top of $r$ left singular vectors of the matrix $\M$.
\\ Initialize: Obtain $\Phat_0$ by $C (\log r)$ iterations of AltProj on $\Y_{[1,t_\train]}$ with $t_\train=C r$ followed by SVD on the output $\Lhat$.
}
}
\label{norst_basic}
\begin{algorithmic}[1]
\algrenewcommand\algorithmicindent{0.5em}
\State \textbf{Input}:   $\y_t$,  \textbf{Output}:  $\shat_t$, $\lhat_t$,  $\Phat_{(t)}$, $\That_t$
\State \textbf{Parameters:} $K \leftarrow C\log(1/\zz)$, $\alpha \leftarrow C r \log n$, $\omega_{supp} \leftarrow \xmint/2$, $\xi \leftarrow \xmint/15$,  $r$ 
\State {\bf Initialize: } $j~\leftarrow~1$, $k~\leftarrow~1$ $\Phat_{t_\train} \leftarrow \Phat_{0}$ 
\For{$t > t_\train$}  \tikzmark{right} \tikzmark{top}
\State {\cred \% Projected CS or Robust Regression}
\State $\tty_t \leftarrow \bpsi \y_t$ where $\bpsi \leftarrow \I - \Phat_{(t-1)}\Phat_{(t-1)}{}'$
\State $\xhat_{t,cs} \leftarrow \arg\min_{\tilde{\x}} \norm{\tilde{\x}}_1 \ \text{s.t.} \ \norm{ \tty_t - \bpsi \tilde{\x} } \leq \xi$.
\State $\That_t \leftarrow \{i:\ |\xhat_{t,cs}| > \omega_{supp} \}$.
\State $\xhat_t \leftarrow \I_{\That_t} ( \bpsi_{\That_t}{}' \bpsi_{\That_t} )^{-1} \bpsi_{\That_t}{}'\tty_t$. \tikzmark{bottom}
\State $\lhat_t \leftarrow \yt - \xhat_t$.  \tikzmark{right2}
\State{\cred \% Subspace Update}
\If {$t = t_j + k \alpha$}   \tikzmark{top2} 
\State $\Phat_{j, k} \leftarrow  \SVD_r[\Lhat_{t; \alpha}]$, $\Phat_{(t)} \leftarrow \Phat_{j,k}$, $k \leftarrow k + 1$
\Else
\State $\Phat_{(t)} \leftarrow \Phat_{(t-1)}$
\EndIf
\If{$t = t_j + K\alpha$}
\State $\Phat_{j} \leftarrow \Phat_{(t)}$, $k \leftarrow 1$, $j \leftarrow j+1$
\EndIf \tikzmark{bottom2}
\EndFor
\end{algorithmic}
\end{algorithm}

\subsection{Recursive Projected Compressive Sensing (ReProCS)}
\label{reprocs_details}

The ReProCS code is at \url{http://www.ece.iastate.edu/~hanguo/PracReProCS.html#Code_} and the simple version (which comes with guarantees) is at \url{https://github.com/praneethmurthy/ReProCS}.

In \cite{rrpcp_allerton,rrpcp_perf}, a novel solution framework called Recursive Projected Compressive Sensing (ReProCS) was introduced to solve the dynamic RPCA and the robust subspace tracking (RST) problems. In later works \cite{rrpcp_isit15,rrpcp_aistats, rrpcp_dynrpca,rrpcp_medrop}, this was shown to be provably correct under progressively weaker assumptions.  
Under two simple extra assumptions (beyond those needed by standard RPCA) , after a coarse initialization computed using PCP or AltProj applied to only the first $C r$ samples, ReProCS-based algorithms provide an online, fast, and very memory-efficient (in fact, nearly memory optimal) tracking solution, that also has significantly improved outlier tolerance compared to all solutions for RPCA. By ``tracking solution'' we mean that it can detect and track a change in subspace within a short delay.
The extra assumptions  are: (i) slow subspace change (or fixed subspace); and (ii) outlier magnitudes are either all large enough, or most are large enough and the rest are very small.  (i) is a natural assumption for static camera videos since they do not involve sudden scene changes. (ii) is also a simple requirement because, by definition, an ``outlier'' is a large magnitude corruption. The small magnitude ones can be clubbed with the small noise $\vt$.

The other assumptions needed by ReProCS for RST are the same as, or similar to, the standard ones needed for RPCA identifiability (described earlier). The union of the column spans of all the $\P_j$'s is equal to the span of the left singular vectors of $\L$. Thus, left incoherence is equivalent to assuming that the $\P_j$'s are $\mu$-incoherent.
We replace the right singular vectors' incoherence by an independent identically distributed (i.i.d.) assumption\footnote{Actually, instead of identically distributed, something weaker suffices: same mean and covariance matrix for all times $t$ is sufficient.} on the $\at$'s, along with element-wise bounded-ness. As explained in \cite{rrpcp_medrop}, the two assumptions are similar; the latter is more suited for RST which involves either frame by frame processing or operations on mini-batches of the full data. 
Moreover, since RST algorithms are online, we also need to re-define $\outfracrow$ as follows.
\begin{definition}
Let $\outfracrow^\alpha$ be the maximum nonzero fraction per row in any $\alpha$-consecutive-column sub-matrix of $\S$. Here $\alpha$ is the mini-batch size used by ReProCS. 
\end{definition}
We use $\outfraccol$ as defined earlier. Using the outlier support size, it is also equal to $\max_t |\T_t|/n$.
%

We describe here the most recently studied ReProCS algorithm, Nearly Optimal RST via ReProCS or ReProCS-NORST \cite{rrpcp_medrop}. This is also the simplest and has the best guarantees. 
It starts with a ``good'' estimate of the initial subspace, which is obtained by $C (\log r)$ iterations of AltProj applied to $\Y_{[1,t_\train]}$ with $t_\train=C r$. It then iterates between (a) {\em Projected Compressive Sensing (approximate Robust Regression)\footnote{As explained in detail in \cite{rrpcp_review}, projected CS is equivalent to solving the robust regression (RR) problem with a sparsity model on the outliers. To understand this simply, let $\Phat= \Phat_{(t-1)}$. The exact version of robust regression assumes that the data vector $\yt$ equals $\Phat \at + \x_t$, while its approximate version assumes that this is only approximately true.  Since $\Phat$ is only an approximation to $\P_{(t)}$, even in the absence of unstructured noise $\vt$, approximate RR is the correct problem to solve for our setting.
Approximate RR solves $\min_{\a,\x} \lambda \|\x\|_1+ \|\yt - \Phat \a - \x\|^2$. In this, one can solve for $\a$ in closed form to get $\hat\a = \Phat'(\yt-\x)$. Substituting this, approximate RR simplifies to $\min_{\x} \|\x\|_1+ \|(\I - \Phat \Phat') (\yt - \x)\|^2$. This is the same as the Lagrangian version of projected CS. The version for which we obtain guarantees solves $\min_{\x} \|\x\|_1 \text{ s.t. } \|(\I - \Phat \Phat') (\yt - \x)\|^2 \le \xi^2$, but we could also have used other CS results and obtained guarantees for the Lagrangian version with minimal changes.}}
in order to estimate the sparse outliers, $\xt$, and then $\lt$ as $\lhat_t = \mt- \xhat_t$; and (b) {\em Subspace Update} to update the subspace estimate $\Phat_{(t)}$. Subspace update is done once every $\alpha$ frames. At each update time, it toggles between the ``detect'' and ``update'' phases. In the detect phase, the current subspace has been accurately updated, and the algorithm is only checking if the subspace has changed. Let $\Phat = \Phat_{(t-1)}$. This is done by checking if the maximum singular value of  $ (\I - \Phat \Phat')[\lhat_{t-\alpha+1},\lhat_{t-\alpha+2},\dots, \lhat_{t}]$ is above a threshold. Suppose the change is detected at $\that_j$. At this time, the algorithm enters the ``update'' phase. This involves obtaining improved estimates of the new subspace by $K$ steps of $r$-SVD, each done with a new set of $\alpha$ samples of $\lhatt$. 
At $t = \that_j + K \alpha$, the algorithm again enters the ``detect'' phase.
We summarize an easy-to-understand version of the algorithm in Algorithm \ref{norst_basic}. 
%


The simplest projected CS (robust regression) step consists of $l_1$ minimization followed by support recovery and LS estimation as in Algorithm \ref{norst_basic}. However, this can be replaced by any other CS solutions, including those that exploit structured sparsity (assuming the outliers have structure).

The guarantee for ReProCS-NORST says the following \cite{rrpcp_medrop}.
To keep the statement simple, the condition number $f$ of $\E[\at \at{}']$ is treated as a numerical constant.

\begin{theorem}
Consider ReProCS-NORST.
Let $\alpha := C  r \log n$,
$\Lam:= \E[\a_1 \a_1{}']$, $\lambda^+:=\lambda_{\max}(\Lam)$, $\lambda^-:=\lambda_{\min}(\Lam)$, 
and let $\xmint:=\min_t \min_{i \in \T_t} (\xt)_i$ denote the minimum outlier magnitude.
Pick an $\zz \leq \min(0.01,0.4 \min_j \SE(\P_{j-1}, \P_j)^2/f)$.
If
\ben
\item $\P_j$'s are $\mu$-incoherent; and $\at$'s are zero mean, mutually independent over time $t$, have identical covariance matrices, i.e. $\E[\at \at{}'] = \Lam$, are
element-wise uncorrelated ($\Lam$ is diagonal), are element-wise bounded (for a numerical constant $\eta$, $(\at)_i^2 \le \eta \lambda_i(\Lam))$, and are independent of all outlier supports $\T_t$,

\item  $\|\vt\|^2 \le c r \|\E[\vt \vt{}']\|$, $\|\E[\vt \vt{}']\| \le c \zz^2 \lambda^-$, $\vt$'s are zero mean, mutually independent, and independent of $\xt,\lt$;

\item $\outfraccol \le  {c_1}/{\mu r}$, $\outfracrow^\alpha \le c_2$, 

\item subspace change: let $\dif:=\max_j \SE(\P_{j-1}, \P_j)$,
\ben
\item $t_{j+1}-t_j > C r \log n \log(1/\zz)$, and
\item $\dif \le 0.8$ and $C_1 \sqrt{r \lambda^+} (\dif + 2\zz) \le \xmint$;
\een

\item initialize\footnote{This can be satisfied by applying $C \log r$ iterations of AltProj \cite{robpca_nonconvex} on the first $C r$ data samples and assuming that these have outlier fractions in any row or column bounded by $c/r$.}:
{$\SE(\Phat_0,\P_0) \le 0.25$, $C_1 \sqrt{r \lambda^+}  \SE(\Phat_0,\P_0) \le \xmint$;}%

\een
and algorithm parameters are appropriately set\footnote{Need knowledge of $\lambda^+$, $\lambda^-$, $r$, $\xmint$.},
then,
\\ with probability $\ge 1 - 10 \tmax n^{-10} $, 
$\SE(\Phat_{(t)}, \P_{(t)}) \le $
\[
 \left\{
\begin{array}{ll}
(\zz + \dif) & \text{ if }  t \in [t_j, \that_j+\alpha), \\
 (0.3)^{k-1} (\zz + \dif) & \text{ if }  t \in [\that_j+(k-1)\alpha, \that_j+ k\alpha), \\
\zz   & \text{ if }  t \in [\that_j+ K\alpha+\alpha, t_{j+1}),
\end{array}
\right.
\]
where $K := C \log (1/\zz)$.
Memory complexity is $O(n r \log n \log (1/\zz) )$ and time complexity is $O(n \tmax r \log (1/\zz) )$.%
\label{thm1}
\end{theorem}
Under Theorem \ref{thm1} assumptions, the following also hold:
\ben
\item {$\|\xhat_t-\xt\| = \|\lhat_t-\lt\| \le 1.2 (\SE(\Phat_{(t)}, \P_{(t)}) + \zz) \|\lt\|$ } with $\SE(\Phat_{(t)}, \P_{(t)})$ bounded as above.

\item  at all times, $t$, $\That_t = \T_t$,

\item $t_j \le \that_j \le t_j+2 \alpha$,

\item Offline-NORST: $\SE(\Phat_{(t)}^{offline}, \P_{(t)})  \le \zz$, $\|\xhat_t^{offline}-\xt\| = \|\lhat_t^{offline}-\lt\| \le \zz \|\lt\|$ at all $t$.
\een



\begin{remark}
The outlier magnitudes lower bound assumption of Theorem \ref{thm1} can be relaxed to a lower bound on most outlier magnitudes. In particular, the following suffices: assume that
$\xt$ can be split into $\xt = (\xt)_{small} + (\xt)_{large}$ that are such that, in the $k$-th subspace update interval,
$\|(\xt)_{small}\|  \le 0.3^{k-1} (\zz+\dif)\sqrt{r \lambda^+} $ and
the smallest nonzero entry of $(\xt)_{large}$ is larger than $C 0.3^{k-1} (\zz+\dif)\sqrt{r \lambda^+} $.
If there were a way to bound the element-wise error of the CS step (instead of the $l_2$ norm error), the above requirement could be relaxed further.
\label{relax_lowerbound}
\end{remark}

A key advantage of ReProCS-NORST is that it automatically detects and tracks subspace changes, and both are done relatively quickly. Theorem \ref{thm1} shows that, with high probability (whp), the subspace change gets detected within a delay of at most $2\alpha = C (r \log n)$ time instants, and the subspace gets estimated to $\zz$ error within at most $(K+2) \alpha = C  (r \log n)\log (1/\zz)$ time instants. Observe that both are nearly optimal since $r$ is the number of samples needed to even specify an $r$-dimensional subspace.
The same is also true for the recovery error of $\xt$ and $\lt$.
If offline processing is allowed, with a delay of at most $C (r \log n)\log (1/\zz)$ samples, we can guarantee all recoveries within normalized error $\zz$.%

Theorem \ref{thm1} also shows that ReproCS-NORST {tolerates a constant maximum fraction of outliers per row (after initialization), without making any assumption on how the outliers are generated.} We explain in Sec. \ref{outrow_explain} why this is possible. This is better than what {\em all} other RPCA solutions allow: all either need this fraction to be $O(1/\rmat)$ or assume that the outlier support is uniform random.
%
The same is true for its memory complexity which is almost $\tmax/r$ times better than all others. 

We should clarify that NORST allows the maximum fraction of outliers per row to be $O(1)$ but this does not necessarily imply that the number of outliers in each row can be this high. The reason is it only allows the fraction per column to only be $O(1/r)$. Thus, for a matrix of size $n \times \alpha$, it allows the total number of outliers to be $O(\min(n\alpha, n\alpha/r)) = O(n\alpha/r)$. Thus the average fraction allowed is only $O(1/r)$.%

ReProCS-NORST has the above advantages only if a few extra assumptions hold. The first is element-wise boundedness of the $\at$'s. This, along with mutual independence and identical covariance matrices of the $\at$'s, is similar to the right incoherence assumption needed by all static RPCA methods. To understand this point, see \cite{rrpcp_medrop}. The zero-mean and diagonal $\Lam$ assumptions are minor. 
The main extra requirement is that $\xmint$ be lower bounded as given in the last two assumptions of Theorem \ref{thm1}.
The lower bound is reasonable as long as the initial subspace estimate is accurate enough and the subspace changes slowly enough so that both $\dif$ and $\SE(\Phat_0,\P_0)$ are $O(1/\sqrt{r})$. This requirement may seem restrictive on first glance but actually is not. The reason is that $\SE(.)$ is only measuring the largest principal angle. This bound still allows the chordal distance between the two subspaces to be $O(1)$. Chordal distance  \cite{chordal_dist}  is the $l_2$ norm of the vector containing the sine of all principal angles. 
This can be satisfied by running just $C \log r$ iterations of AltProj on a short initial dataset: just $t_\train=C r$ frames suffice.%

{\em Why ReProCS works. }
Let $\bpsi :=(\I - \Phat_{(t-1)} \Phat_{(t-1)}{}')$. As also briefly explained in \cite{rrpcp_review, rrpcp_medrop}, it is not hard to see that the ``noise'' $\b_t:=\bm\Psi \lt$ seen by the projected CS step is proportional to the error between the subspace estimate from $(t-1)$ and the current subspace.
Moreover, incoherence (denseness) of the $\P_{(t)}$'s and slow subspace change together imply that $\bpsi$ satisfies the restricted isometry property (RIP) \cite{rrpcp_perf}.
Using these two facts, a result for noisy $l_1$ minimization, and the lower bound assumption on outlier magnitudes, one can ensure that the CS step output is accurate enough and the outlier support $\T_t$ is correctly recovered. With this, it is not hard to see that $\lhat_t = \lt + \vt - \et$ where $\et:=\x_t -\xhat_t$ satisfies
\[
\et= \I_{\T_t} (\bm\Psi_{\T_t}{}'\bm\Psi_{\T_t})^{-1} \I_{\T_t}{}' \bm\Psi' \lt
\]
and $\|\et\| \le C \|\b_t\|$. 
Consider subspace update. Every time the subspace changes, one can show that the change can be detected within a short delay. After that, the $K$ SVD steps help get progressively improved estimates of the changed subspace. To understand this, observe that, after a subspace change, but before the first update step, $\b_t$ is the largest and hence, $\et$, is also the largest for this interval. However, because of good initialization or because of slow subspace change and previous subspace correctly recovered (to error $\zz$), neither is too large. Both are proportional to $(\zz + \dif)$, or to the initialization error. Recall that $\dif$ quantifies the amount of subspace change. For simplicity suppose that $\SE(\Phat_0,\P_0) = \dif$.
Using the idea below, we can show that we get a ``good'' first estimate of the changed subspace.

The input to the PCA step is $\lhat_t$ and the noise seen by it is $\et$. Notice that $\et$ depends on the true data $\lt$ and hence this is a setting of PCA in data-dependent noise \cite{corpca_nips,pca_dd}. From \cite{pca_dd}, it is known that the subspace recovery error of the PCA step is proportional to the ratio between the time-averaged noise power plus time-averaged signal-noise correlation, $(\|\sum_t \E[\et \et{}']\| + \|\sum_t \E[\lt \et{}'\|)/\alpha$, and the minimum signal space eigenvalue, $\lambda^-$.
The instantaneous values of both noise power and signal-noise correlation are of order $(\dif+\zz)$ times $\lambda^+$. However, using the fact that $\et$ is sparse with support $\T_t$ that changes enough over time so that $\outfracrow^\alpha$ is bounded, their time-averaged values are at least $\sqrt{\outfracrow^\alpha}$ times smaller. This follows using Cauchy-Schwartz. As a result, after the first subspace update, the subspace recovery error is below $4 \sqrt{\outfracrow} (\lambda^+/\lambda^-)$ times $(\dif+\zz)$. Since $\outfracrow (\lambda^+/\lambda^-)^2$ is bounded by a constant $c_2 < 1$, this means that, after the first subspace update, the subspace error is below $\sqrt{c_2}$ times $(\dif+\zz)$.

This, in turn, implies that $\|\b_t\|$, and hence $\|\et\|$, is also $\sqrt{c_2}$ times smaller in the second subspace update interval compared to the first. This, along with repeating the above argument, helps show that the second estimate of the changed subspace is $\sqrt{c_2}$ times better than the first and hence its error is $(\sqrt{c_2})^2$ times $(\dif+\zz)$. Repeating the argument $K$ times, the $K$-th estimate has error $(\sqrt{c_2})^K$ times $(\dif+\zz)$. Since $K = C \log(1/\zz)$, this is an $\zz$ accurate estimate of the changed subspace.%

\subsection{Looser bound on $\outfracrow$ and outlier magnitudes' lower bound}\label{outrow_explain}
As noted in \cite{robpca_nonconvex}, solutions for standard RPCA (that only assumes left and right incoherence of $\L$ and nothing else) cannot tolerate a bound on outlier fractions in any row or any column that is larger than $1/\rmat$\footnote{The reason is this: let $b_0 = \outfracrow$, one can construct a matrix $\X$ with $b_0$ outliers in some rows that has rank equal to $1/b_0$. A simple way to do this would be to let the support and nonzero entries of $\X$ be constant for $b_0 \tmax$ columns before letting either of them change. Then the rank of $\X$ will be $\tmax/(b_0 \tmax)$. A similar argument can be used for $\outfraccol$.}.
The reason ReProCS-NORST can tolerate a constant $\outfracrow^\alpha$ bound is because it uses extra assumptions. We explain the need for these here (also see \cite{rrpcp_review,rrpcp_medrop} for a brief version of this explanation). ReProCS recovers $\xt$ first and then $\lt$ and does this at each time $t$.
When recovering $\xt$, it exploits ``good'' knowledge of the subspace of $\lt$ (either from initialization or from the previous subspace's estimate and slow subspace change), but it has no way to deal with the residual error, $\b_t:= (\I - \Phat_{(t-1)} \Phat_{(t-1)}{}') \lt$, in this knowledge. Since the individual vector $\b_t$ does not have any structure that can be exploited\footnote{However the $\b_t$'s arranged into a matrix do form a low-rank matrix whose approximate rank is $r$ or even lower (if not all directions change). If we try to exploit this structure we end up with the modified-PCP type approach. This either needs the uniform random support assumption (used in its current guarantee) or even if a different proof approach is used, for identifiability reasons similar to the one described above, it will still not tolerate outlier fractions larger than $1/r_\new$ where $r_\new$ is the (approximate) rank of the matrix formed by the $\b_t$'s.}, the error in recovering $\xt$ cannot be lower than $C \|\b_t\|$. This means that, to correctly recover the support of $\xt$, $\xmint$ needs to be larger than $C\|\b_t\|$. This is where the $\xmint$ lower bound comes from. If there were a way to bound the element-wise error of the CS step (instead of the $l_2$ norm error), we could relax the $\xmint$ bound significantly. 
Correct support recovery is needed to ensure that the subspace estimate can be improved with each update. In particular, it helps ensure that the error vectors $\et:=\xt - \xhat_t$ in a given subspace update interval are mutually independent when conditioned on the $\yt$'s from all past intervals.
This step also uses element-wise boundedness of the $\at$'s along with their mutual independence and identical covariances. As noted earlier, these replace the right incoherence assumption.

\subsection{Simple-ReProCS and older ReProCS-based solutions}

The above result for ReProCS-NORST is the best one \cite{rrpcp_medrop,rrpcp_merop}. It improves upon our recent guarantee for simple-ReProCS \cite{rrpcp_dynrpca}.
The first part of the simple-ReProCS algorithm (robust regression step) is the same as ReProCS-NORST. The subspace update step is different.  After a subspace change is detected, this involves $K$ steps of projection-SVD  or ``projection-PCA'' \cite{rrpcp_perf}, each done with a new set of $\alpha$ frames of $\lhatt$; followed by an $r$-SVD based subspace re-estimation step, done with another new set of $\alpha$ frames. The projection-SVD steps are less expensive since they involve a 1-SVD instead of an $r$-SVD, thus making simple-ReProCS faster. It has the following guarantee \cite{rrpcp_dynrpca}.
\begin{theorem}
Consider simple-ReProCS \cite{rrpcp_dynrpca}. If
\bi
\item first  three assumptions of Theorem \ref{thm1} holds,

\item subspace change: assume that 
only one subspace direction changes at each $t_j$, and
 $C \sqrt{\lambda_\ch} \dif + 2\sqrt{\lambda^+}\zz \le \xmint$, where  $\dif:=\max_j \SE(\P_{j-1}, \P_j)$ and $\lambda_\ch$ is the eigenvalue along the changing direction,

\item init:
 $\SE(\Phat_0,\P_0) \le \zz$,
\ei
Then all conclusions of Theorem \ref{thm1} hold.
\end{theorem}
 Simple-ReProCS shares most of the advantages of ReProCS-NORST. Its disadvantage is that it requires that, at each change time, only one subspace direction changes. Because of this, even though its tracking delay is the same as that of ReProCS-NORST, it is $r$-times sub-optimal. Moreover, it needs the initial subspace estimate to be $\zz$-accurate.

The above two guarantees are both a significant improvement upon the earlier partial \cite{rrpcp_perf} and complete \cite{rrpcp_isit15,rrpcp_aistats} guarantees for original-ReProCS. These required a very specific model on outlier support change (instead of just a bound on outlier fractions per row and per column); needed an unrealistic model of subspace change and required the eigenvalues along newly added directions to be small for some time.

\input{tables_proc_ieee}

\subsection{Heuristics for Online RPCA and/or RST}
Almost all existing literature other than the modified-PCP and ReProCS frameworks described above focus on incremental, online, or streaming solutions for RPCA. Of course any online or incremental RPCA solution will automatically also provide a tracking solution if the underlying subspace is time-varying. Thus, algorithmically, incremental RPCA, online RPCA, or tracking algorithms are the same. The performance metrics for each case are different though. All of these approaches come with either no guarantee or a partial guarantee (the guarantee depends on intermediate algorithm estimates).

Early attempts to develop incremental solutions to RPCA that did not explicitly use the S+LR definition include \cite{Li03anintegrated,ipca_weightedand}. 
%
The first online heuristic for the S+LR formulation was called Real-time Robust Principal Components Pursuit (RRPCP) \cite{rrpcp_allerton}. The algorithm name is a misnomer though since the method had nothing to do with PCP which requires solving a convex program. In fact, RRPCP was a precursor to the ReProCS framework \cite{rrpcp_allerton11,rrpcp_tsp,rrpcp_perf}  described above. The first guarantee for a ReProCS algorithm was proved in \cite{rrpcp_perf}. This was a partial guarantee though (it assumed that intermediate algorithm estimates satisfy certain properties). However, the new proof techniques introduced in this work form the basis of all the later complete guarantees including the ones described above.
An online solution that followed soon after was ORPCA \cite{xu_nips2013_1}. This is an online stochastic optimization based solver for the PCP convex program. This also came with only a partial guarantee (the guarantee assumed that the subspace estimate outputted at each time is full rank).
%
%
Approaches that followed up on the basic ReProCS idea of alternating the approximate Robust Regression (projected Compressive Sensing) and subspace update steps include GRASTA \cite{grass_undersampled}, pROST \cite{seidel2013prost} and ROSETA \cite{mansour_robust_ss_track}. GRASTA replaced both the steps by different and approximate versions.
 It solves the exact version of robust regression which involves recovering $\at$ as $\arg\min_{\a} \|\yt - \Phat_{t-1} \a\|_1$. This approach ignores the fact that $\Phat_{(t-1)}$ is only an approximation to the current subspace $\P_{(t)}$. This is why, in experiments, GRASTA fails when there are significant subspace changes: it ends up interpreting the subspace tracking error as an outlier.  
In its subspace update step, the SVD or projected-SVD used in different variants of ReProCS \cite{rrpcp_allerton11,rrpcp_tsp,rrpcp_medrop} are replaced by a faster but approximate subspace tracking algorithm called GROUSE \cite{grouse} that relies on stochastic gradient descent. 
Both of pROST and ROSETA modify the GRASTA approach, and hence, indirectly rely on the basic ReProCS framework of alternating robust regression and subspace update.  pROST replaces $l_1$ minimization by non-convex $l_0$-surrogates. In ROSETA, an ADMM algorithm is used to solve the robust regression.

\section{Pros and Cons of Various Approaches}
We provide a summary of the comparisons in Table \ref{compare_assu}. 
We discuss our main conclusions here.
\ben

\item {\em Outlier tolerance. } The PCP (C) and the modified-PCP results allow the loosest upper bounds on  $\outfracrow$ and $\outfraccol$, however both allow this only under a uniform random support model. This is a restrictive assumption. For example, for the video application, it requires that video objects are only one or a few pixels wide and jumping around randomly. AltProj, GD, NO-RMC, and ReProCS do not assume any outlier support model. Out of these, GD needs $\max(\outfracrow,\outfraccol) \in O(1/\rmat^{1.5})$, AltProj and NO-RMC only need $\max(\outfracrow,\outfraccol) \in O(1/\rmat)$, while ReProCS-NORST has the best outlier tolerance of $\outfracrow^\alpha \in O(1)$ and $\outfraccol \in O(1/r)$.  For the video application, this means that it allows large-sized and/or slow-moving or occasionally static foreground objects much better than all other approaches. Also see Sec. \ref{sims}.

\item {\em Nearly square data matrix. } Only NO-RMC needs this. This can be an unreasonable requirement for videos which often have much fewer frames $\tmax$ than the image size $n$. NO-RMC needs this because it is actually a robust matrix completion solution; to solve RPCA, it deliberately undersamples the entire data matrix $\Y$ to get a faster RPCA algorithm. The undersampling necessitates a nearly square matrix.

\item {\em Lower bound on most outlier magnitudes. } Only ReProCS requires this extra assumption. This requirement is encoding the fact that outliers are large magnitude corruptions; the small magnitude ones get classified as the unstructured noise $\vt$. As explained earlier in Sec. \ref{outrow_explain}, ReProCS needs this because it is an online solution that recovers $\xt$'s and their support sets $\T_t$, and $\lt$'s on a frame by frame basis and updates the subspace once every $\alpha = C r\log n$ frames.

\item {\em Slow subspace change or fixed subspace. } Both ReProCS and modified-PCP need this. The modified-PCP requirement is often unrealistic, while that of ReProCS-NORST is simple. It should hold for most static camera videos (videos with no scene changes).

\item {\em Incoherence. } All solutions need a form of left and right incoherence. ReProCS replaces the traditional right incoherence assumption with a statistical model on the $\at$'s. This is needed because it is an online solution that updates the subspace once every $\alpha = C r\log n$ frames using just these many past frames. These help ensure that each update improves the subspace estimate.
\een

Speed, memory and other features are as follows.
\ben
\item  {\em Memory complexity. } ReProCS-NORST has the best memory complexity that is also nearly optimal. All other static RPCA solutions need to store the entire matrix.

\item {\em Speed. } NO-RMC is the fastest, ReProCS is the second fastest. Both need extra assumptions discussed above. AltProj is the fastest solution that does not need any extra assumptions.

\item {\em Algorithm parameters. } PCP is the only approach that needs just one algorithm parameter $\lambda$ and the PCP (Cand{\`e}s et al) result is the only one that does not assume any model parameter knowledge to set this parameter. Of course PCP is a convex program which needs a solver; the solver itself does have other parameters to set.

\item  {\em Detecting and tracking change in subspace. } Only ReProCS can do this; ReProCS-NORST is able to do this with near-optimal delay.

\een

\section{Static and Dynamic Matrix Completion}

\subsection{Matrix Completion}
\subsubsection{Nuclear norm minimization}
The first solution to matrix completion was nuclear norm minimization (NNM) \cite{matcomp_first}. This was a precursor to PCP and works for the same reasons that PCP works (see Sec. \ref{pcp_section}). The first guarantee for NNM appeared in \cite{matcomp_candes}.
This result was improved and its proof simplified in \cite{gross,recht_mc_simple}.

\subsubsection{Faster solutions: alternating minimization and gradient descent}
Like PCP, NNM is slow. To address this, alternating minimization and gradient descent solutions were developed in \cite{optspace} and \cite{lowrank_altmin} (and many later works) along with a carefully designed spectral initialization that works as follows: compute $\hat\LU^{0,temp}$ as the matrix of top $r$ singular vectors of $\Y$ (recall $\Y:=\pP_\Omega(\L)$) and then compute $\hat\LU^0$ as its ``clipped'' version as follows: zero out all entries of $\hat\LU^{0,temp}$ that have magnitude more than $2\mu \sqrt{r/n}$ followed by orthonormalizing the resulting matrix. Here clipping is used to ensure incoherence holds for the initialization of $\LU$.
After this, the alt-min solution  \cite{lowrank_altmin} alternatively minimizes $||\pP_{\Omega_{2j+1}}(\L) - \pP_{\Omega_{2j+1}}( \tilde\LU \tilde\LV')||_F^2$ over $\tilde\LU, \tilde\LV$. With keeping one of them fixed, this is clearly a least squares problem which can be solved efficiently. Here, $\Omega_j$ is an independently sampled set of entries from the original observed set $\Omega$. Thus, each iteration uses a different set of samples. This solution has the following guarantee \cite[Theorem 2.5]{lowrank_altmin}:
\begin{theorem}
Let $\L \svdeq \LU \bm\Sigma \LV'$ be its reduced SVD. Consider the alt-min solution \cite{lowrank_altmin}.
If
\ben
\item $\LU$ is $\mu$-incoherent, $\LV$ is $\mu$-incoherent,
\item alt-min is run for $T = C \log (||\L||_F/ \epsilon)$ iterations,
\item the $2T+1$ sets $\Omega_j$ are generated independently from $\Omega$
\item each entry of $\Omega$ is generated iid with probability $p$ that satisfies
$p n \tmax \ge C \kappa^3 \mu^2 n r^{3.5} \log n \log (||\L||_F/ \epsilon)$
\een
then, with probability at least $ 1 - c n^{-3}$, $||\L - \hat\LU^T \hat\LV^T{}'||_F \le \epsilon$.
\label{altmin_thm}
\end{theorem}
In the above result, $\kappa$ is the condition number of $\L$. Later works, e.g., \cite{lowrank_altmin_no_kappa}, removed the dependence on condition number by studying a modification of simple alt-min that included a soft deflation scheme that relied on the following intuition: if $\kappa$ is large, then for small ranks $r$,  the spectrum of $\L$ must come in well-separated clusters with each cluster having small condition number. This idea is similar to that of the AltProj solution described earlier for RPCA \cite{robpca_nonconvex}.

\subsubsection{Eliminating the partitioned sampling requirement of previous alt-min results}
The main limitation of the above results for alt-min is the need for the partitioned sampling scheme. This does not use all samples at each iteration (inefficient) and, as explained in \cite[I-B]{mc_luo}, the sampling of $\Omega_j$ is hard to implement in practice.
The above limitation was removed in a the nice recent work of Sun and Luo \cite{mc_luo}. However, this work brought back the dependence on condition number. Thus, their result is essentially similar to Theorem \ref{altmin_thm}, but without the third condition (it uses all samples, i.e., $\Omega_j = \Omega$), and hence, without a dependence of the sample complexity on $\epsilon$. On the other hand, its disadvantage is that it does not specify the required number of iterations, $T$, that above results do, and thus does not provide a bound on computational complexity.
Theorem 3.1 of \cite{mc_luo} says the following:
\begin{theorem}
Let $\L \svdeq \LU \bm\Sigma \LV'$ be its reduced SVD.
Consider either the alt-min or the gradient descent solutions with a minor modification of the spectral initialization idea described above (any of Algorithms 1-4 of \cite{mc_luo}).
If
\ben
\item $\LU$ is $\mu$-incoherent, $\LV$ is $\mu$-incoherent,
\item the set $\Omega$ is generated uniformly at random with size $m$ that satisfies
$m \ge C \kappa^2 \mu^2 n r \max(\mu \log n, \kappa^4 \mu^2 r^6)$
\een
then, with probability at least $ 1 - c n^{-3}$, $||\L - \hat\LU^T \hat\LV^T{}'||_F \le \epsilon$ (the number of iterations required, $T$, is not specified).
\label{altmin_thm}
\end{theorem}

\subsubsection{No bad local minima or saddle points}
Finally, all the above results rely on a carefully designed spectral initialization scheme followed by a particular iterative algorithm. Careful initialization is needed under the implicit assumption that, for the cost function to be minimized, there are many local minima  and/or saddle points and the value of the cost function at the local minima does not equal the global minimum value.
However, a series of very interesting recent works \cite{ge_1,ge_best} has shown that this is, in fact, not the case: even though the cost function is non-convex, all its local minima are global minima and all its saddle points (points at which the gradient is zero but the Hessian matrix is indefinite) have Hessians with at least one negative eigenvalue. For the best result, see Theorems 1, 4 of \cite{ge_best}. In fact, there has been a series of recent works showing similar results also for matrix sensing, robust PCA, and phase retrieval \cite{sun_wright, ge_1,ge_best}. 

%
%

\subsection{Dynamic MC or Subspace Tracking with missing data} \label{subtrack}

In the literature, there are three well-known algorithms for subspace tracking with missing data (equivalently, dynamic MC): PAST \cite{past,past_conv}, PETRELS \cite{petrels} and GROUSE \cite{grouse,local_conv_grouse, grouse_global, grouse_enh}. All are motivated by stochastic gradient descent to solve the PCA problem and the Oja algorithm.  These and many others are described in detail in a review article on subspace \cite{chi_review}. In this section, we briefly summarize the theoretical guarantees this problem: the only result for missing data is for GROUSE for the case of {\em single unknown subspace}. Moreover, the result is a {\em partial guarantee (result makes assumptions on intermediate algorithm estimates)}. We give it next \cite{local_conv_grouse}.

\begin{theorem}[GROUSE for subspace tracking with missing data]
Suppose the unknown subspace is fixed, denote it by $\P$. Let $\epsilon_t := \sum_{i=1}^r \sin^2\theta_i(\Phat_{(t)}, \P)$ where $\theta_i$ is the $i$-th largest principal angle between the two subspaces. Also, for a vector $\bm{z} \in \mathbb{R}^n$, let $\mu(\bm{z}):= \frac{n \|\bm{z}\|_{\infty}^2}{\|\bm{z}\|_{2}^2}$ quantify its denseness.

Assume that (i) the subspace is fixed and denoted by $\P$; (ii) $\P$ is $\mu$-incoherent; (iii) the coefficients vector $\at$ is independently from a standard Gaussian distribution, i.e., $(\at)_i \overset{i.i.d.}{\sim} \mathcal{N}(0, 1)$; (iv) the size of the set of observed entries at time $t$, $\Omega_t$, satisfies $|\Omega_t| \geq  (64/3) r (\log^2 n) \mu \log(20r)$; and the following assumptions on intermediate algorithm estimates hold:
\bi
\item $\epsilon_t \leq \min(\frac{r \mu}{16n}, \frac{q^2}{128 n^2  r} )$;
\item the residual at each time, $\bm{r}_t := \lt - \Phat_{(t)}\Phat_{(t)}' \lt$ satisfies
\begin{align*}
\mu(\bm{r}_t) \leq \min\bigg\{ \log n \left[\frac{0.045}{\log 10} C_1 r \mu \log(20 r)\right]^{0.5}, \\
 \log^2 n \frac{0.05}{8 \log 10} C_1 \log(20 r) \bigg\}
\end{align*}
with probability at least $1 - \bar{\delta}$ where $\bar{\delta} \leq 0.6$.
\ei
Then,
\[
\E[\epsilon_{t+1} | \epsilon_t] \leq \epsilon_t -.32(.6 - \bar{\delta}) \frac{q}{nr}\epsilon_t + 55 \sqrt{\frac{n}{q}} \epsilon_t^{1.5}.
\]
\end{theorem}
The above result is a partial guarantee and is hard to parse. Its denseness requirement on the residual $\bm{r}_t$ is reminiscent of the denseness assumption on the currently un-updated subspace needed by the first (partial) guarantee for ReProCS from \cite{rrpcp_perf}.

The only complete guarantee for subspace tracking exists for the case of no missing data \cite{grouse_global}. It still assumes a {\em single unknown} subspace. We give this next.

\begin{theorem}[GROUSE-full]
Given data vectors $\yt = \lt$ (no noise and no missing data).
Suppose the unknown subspace is fixed, denote it by $\P$. Let $\Phat_{(t)}$ denote the estimate of the subspace at time $t$. Let $\epsilon_t = \sum_{i=1}^r \sin^2\theta_i(\Phat_{(t)}, \P)$ as before.
Assume that the initial estimate, $\Phat_{(0)}$ is obtained by a random initialization, i.e., it is obtained by orthonormalizing a $n \times r$ i.i.d. standard normal matrix. Then, for any $\epsilon_\ast > 0$ and any $\delta, \delta{}' >0$, after
$
T= K_1 + K_2 = \left(\frac{r^3}{\rho{}'} + r \right)\mu_0 \log n + 2r\log\left( \frac{1}{\epsilon_\ast \rho}\right)
$
iterations, with probability at least $1 - \delta - \delta{}'$, GROUSE satisfies $\epsilon_T \leq \epsilon_\ast$. Here, $\mu = 1 + \frac{\log((1 - \delta{}')/C + r \log(e/r)}{r\log n}$ for a constant $C \approx 1$.

In the noisy but no missing data setting, i.e., when $\yt = \lt + \vt$, the following can be claimed for GROUSE:
$
\mathbb{E}\left[\epsilon_{t+1} \big| \Phat_{(t)} \right] \le \left( 1 - \frac{\beta_0}{r}\left(\cos^2\theta_1(\Phat_{(t)}, \P) - \frac{\beta_1\sigma^2}{\epsilon_t/r + \beta_1\sigma^2}\right)\right)\epsilon_t
$
where $\beta_0 = \frac{1}{1 + r\sigma^2/n}$, $\beta_1 = 1 - r/n$.
\end{theorem}

\input{proc_ieee_expts_3}
\section{Conclusions and Future Directions}

The original or static RPCA problem as well as the MC problem have been extensively studied in the last decade. However robust subspace tracking (RST) has not received much attention until much more recently. The same is also true in terms of provably correct solutions for subspace tracking with missing data (or dynamic MC) as well.
In \cite{rrpcp_merop,rrpcp_medrop} a simple and provably correct RST approach was obtained that works with near-optimal tracking delay under simple assumptions - weakened versions of standard RPCA assumptions, lower bound on outlier magnitudes, true data subspace either fixed or slowly changing, and a good initialization  (obtained via few iterations of AltProj applied to the first $C r$ data samples). After initialization, it can tolerate a constant bound on maximum outlier fractions in any row of later mini-batches of the data matrix. This is better than what any other RPCA solution can tolerate. But, this is possible only because of the extra assumptions. As explained earlier, the only way to relax the outlier magnitudes lower bound is if one could bound the element-wise error of the Compressive Sensing step.

\subsubsection{Subspace Tracking: dynamic RPCA, MC, RMC, and undersampled RPCA}
Consider dynamic RPCA or robust subspace tracking (RST). Two tasks for future work are (a) replace the projected CS / robust regression step which currently uses simple $l_1$ minimization by more powerful CS techniques such as those that exploit structured sparsity; and (b) replace the SVD or projected-SVD in the subspace update step by fully streaming (single pass) algorithms. 
Both have been attempted in the past, but without provable guarantees. In the algorithm developed and evaluated for videos in \cite{rrpcp_tsp,rrpcp_isit},  slow (or model-driven) support change of the foreground objects was exploited. The GRASTA approach \cite{grass_undersampled} used a stochastic gradient descent approach called GROUSE for subspace update.
Two more difficult open questions include: (a) provably dealing with moving cameras (or more generally with small group transformations applied to the observed data), and (b) being able to at least detect sudden subspace change while being robust to outliers. For the video application, this would occur due to sudden scene changes (camera being turned around for example).
In the ReProCS approaches studied so far, a sudden subspace change would get confused as a very large support outlier where as in the subspace tracking with missing data approaches such as GROUSE or PETRELS, all guarantees are for the case of a fixed unknown subspace. Some heuristics for dealing with moving cameras include \cite{video_app1, video_app2}.

Two important extensions of the RST problem include dynamic robust matrix completion (RMC) or RST with missing data and undersampled RST. The latter finds applications in undersampled dynamic MRI. There is an algorithm and a partial guarantee for  undersampled RST in \cite{rrpcp_allerton11,rrpcp_tsp,rrpcp_globalsip}, but a complete correctness result still does not exist; and careful experimental evaluations on real dynamic MRI datasets are missing too.
Dynamic RMC finds applications in recommendation system design when the factors governing user preferences can change over time, e.g., as new demographics of users get added, or as more content gets added.
%
In the setting where the number of users is fixed but content gets added with time, we can let $\yt$ be the vector of ratings of content (e.g., movie) $t$ by all users. This vector will have zeros (corresponding to missing entries) and outliers (due to typographical errors or users' laziness). If the content is fixed but the number of users are allowed to increase over time, $\yt$ can be the vector of ratings by the $t$-th user. Either of these cases can be dealt with by initially using AltProj on an initial batch dataset. As more users get added, new ``directions'' will get added to the  factors' subspace. This can be detected and tracked using an RST solution.
An open question is how to deal with the most practical setting when both the users and content is allowed to get added with time.

\subsubsection{Phaseless Robust PCA and Subspace Tracking}
A completely open question is whether one can solve the phaseless robust PCA or S+LR problem. In applications, such as ptychography, sub-diffraction imaging or astronomy, one can only acquire magnitude-only measurements \cite{candes_phaselift}. If the unknown signal or image sequence is well modeled as sparse + low-rank, can this modeling be exploited to recover it from under-sampled phaseless measurements? Two precursors -- low rank phase retrieval \cite{lrpr_tsp} and phase retrieval for a single outlier-corrupted signal \cite{chi_pr_outliers} -- have been recently studied.%

\subsubsection{(Robust) subspace clustering and its dynamic extension}
An open question is how can robust and dynamic robust PCA ideas be successfully adapted to solve other more general related problems. One such problem is subspace clustering \cite{ss_clust} which involves clustering a given dataset into one of $K$ different low-dimensional subspaces.  This can be interpreted as a generalization of PCA which tries to represent a given dataset using a single low-dimensional subspace. Subspace clustering instead uses a union of subspaces to represent the true data, i.e, each data vector is assumed to be generated from one of $K$ possible subspaces.
An important question of interest is the following: given that subspace clusters have been computed for a given dataset, if more data vectors come in sequentially, how can one incrementally solve the clustering problem, i.e., either classify the new vector into one of the $K$ subspaces, or decide that it belongs to a new subspace? Also, under what assumptions can one solve this problem if the data were also corrupted by additive sparse outliers?


We should point out that RST should not be confused as a special case of robust subspace clustering, since, to our best knowledge, robust subspace clustering solutions do not deal with additive sparse outliers. They only deal with the case where an entire data vector is either an outlier or not. Moreover, all subspace clustering solutions require that the $K$ subspaces  be ``different'' enough, while RST requires the opposite.

\bibliographystyle{IEEEbib} 
\bibliography{../../bib/tipnewpfmt_kfcsfullpap,addTB,./SPMAG_expts/bare_jrnl}

\end{document}

%% file: tikz_def_proc.tex
\usepackage{tikz}
\usepackage{pgfplots}
\usepgfplotslibrary{groupplots}
\newcommand{\tikzmark}[1]{\tikz[overlay,remember picture] \node (#1) {};}

\pgfplotsset{
        my stylecompare/.style={
			width=.4\textwidth,
            height=4.5cm,
            label style={font=\Large},
            title style={font=\Large},
            x tick label style={font =\small, /pgf/number format/1000 sep=},
	    axis lines=left,
	    major x tick style = transparent,
	    major y tick style = transparent,
	    every y tick label/.style={
 		   xshift=-.7cm, yshift=-2pt,anchor=south west,inner sep=0pt,font=\small,
 		   scaled y ticks=false,
	    },
	    every x tick label/.style={font = \small},
            scaled x ticks=false,
        },
        my legend style compare/.style={
            legend entries={
            		GRASTA ($0.6$),
            		ORPCA ($7.6$),
            		NORST ($1.0$),	
            		Offline-NORST ($1.7$),
            		Alt Proj ($70.8$),
            		RPCA-GD ($92.2$),
            },
            legend style={
                at={(0.1,1.5)},
                anchor=north west,
            },
            legend columns=2,
	    legend style={font=\tiny},
        },
        cycle multi list={
        {blue, line width=0.6pt, mark=o,mark size=3pt},
        {black, line width=0.6pt, mark=square,mark size=2.5pt},
        {teal, line width=0.6pt, mark=oplus,mark size=2.5pt},
        {red, line width=0.6pt, mark=triangle,mark size=2.5pt},
        {red, solid, line width=0.5pt, mark=oplus,mark size=2.8pt},
        {olive, line width=0.6pt, mark=10-pointed star,mark size=2.5pt},
        {cyan, line width=0.6pt, mark=Mercedes star,mark size=3pt},
        {teal, line width=0.6pt, mark=oplus,mark size=2.5pt},
		},
}

\usetikzlibrary{decorations.markings}
\usetikzlibrary{decorations.pathreplacing}
\def\MarkLt{4pt}
\def\MarkSep{2pt}

\tikzset{
  TwoMarks/.style={
    postaction={decorate,
      decoration={
        markings,
        mark=at position #1 with
          {
              \begin{scope}[xslant=0.2]
              \draw[line width=\MarkSep,white,-] (0pt,-\MarkLt) -- (0pt,\MarkLt) ;
              \draw[-] (-0.5*\MarkSep,-\MarkLt) -- (-0.5*\MarkSep,\MarkLt) ;
              \draw[-] (0.5*\MarkSep,-\MarkLt) -- (0.5*\MarkSep,\MarkLt) ;
              \end{scope}
          }
       }
    }
  },
  TwoMarks/.default={0.5},
}

\usetikzlibrary{positioning}

\usetikzlibrary{calc}

\tikzstyle{block}  = [rectangle, draw, rounded corners, text width=5cm, text centered, minimum height=1em]
\tikzstyle{smallblock}  = [rectangle, draw, rounded corners,text width=1.8cm, text centered, minimum height=1em]
\tikzstyle{input}  = [rectangle, draw, text width=1.2cm, text centered, minimum height=1em]
\tikzstyle{output}  = [rectangle, draw, text width=1.2cm, text centered, minimum height=1em]
\tikzstyle{block1}  = [rectangle, draw, rounded corners,text width=5cm, text centered, minimum height=1em]
\tikzstyle{blockl1}  = [rectangle, draw, rounded corners,text width=4cm, text centered, minimum height=1em]

%% file: Tex_Commands.tex
\newcommand{\norm}[1]{\left\|#1\right\|}

\newtheorem{theorem}{Theorem}

\newtheorem{definition}[theorem]{Definition}
\newtheorem{remark}[theorem]{Remark}

\newcommand{\bi}{\begin{itemize}}
\newcommand{\ei}{\end{itemize}}
\newcommand{\ben}{\begin{enumerate}}
\newcommand{\een}{\end{enumerate}}

\newcommand{\bean}{\begin{eqnarray*} }
\newcommand{\eean}{\end{eqnarray*} }
\newcommand{\bea}{\begin{eqnarray} }
\newcommand{\eea}{\end{eqnarray} }
\newcommand{\ba}{\begin{align*} }
\newcommand{\ea}{\end{align*} }

\newcommand{\nn}{\nonumber}

\newcommand{\dif}{{\text{dif}}}

\newcommand{\bl}{\begin{frame}}
\newcommand{\el} {\end{frame}}

\newcommand{\cred}{\color{red}} 

\newcommand{\svdeq}{\overset{\mathrm{SVD}}=}

\newcommand{\supp}{\text{support}}

\newcommand{\tmax}{d} 

\newcommand{\mt}{\bm{m}_t}
\newcommand{\xt}{\bm{s}_t}

\newcommand{\s}{\bm{s}}

\newcommand{\x}{\bm{s}}
\newcommand{\xhat}{\hat{\x}}

\renewcommand{\l}{\bm{\ell}}
\newcommand{\lt}{\bm{\ell}_t}
\newcommand{\lhat}{\hat{\bm{\ell}}}

\newcommand{\lhatt}{\hat{\l}_t}   
\newcommand{\y}{\bm{m}}
\newcommand{\yt}{\y_t}
\newcommand{\tty}{{\tilde{\y}}}

\renewcommand{\v}{\bm{w}}
\newcommand{\vt}{\v_t}
\newcommand{\V}{\bm{W}}

\renewcommand{\a}{\bm{a}}

\newcommand{\et}{\bm{e}_t}

\newcommand{\new}{\mathrm{new}}

\newcommand{\at}{\bm{a}_t}

\newcommand{\I}{\bm{I}}

\newcommand{\Lam}{\bm{\Lambda}}

\newcommand{\T}{\mathcal{T}}

\newcommand{\pP}{\bm{\mathcal{P}}}

\newcommand{\A}{\bm{A}}

\newcommand{\Shat}{\hat{\bm{S}}}
\newcommand{\shat}{\hat{\bm{s}}}
\newcommand{\Lhat}{\hat{\bm{L}}}

\renewcommand{\P}{\bm{P}}
\newcommand{\LU}{\bm{U}}
\newcommand{\LV}{\bm{V}}

\newcommand{\ch}{{\mathrm{ch}}}

\newcommand{\G}{{\bm{G}}}

\newcommand{\tU}{\tilde{\bm{U}}}
\newcommand{\tV}{\tilde{\bm{V}}}

\renewcommand{\b}{\bm{b}}
\newcommand{\Phat}{\hat{\bm{P}}}

\newcommand{\Span}{\operatorname{span}} 

\newcommand{\basis}{\operatorname{basis}}

\newcommand{\E}{\mathbb{E}}

\newcommand{\train}{\mathrm{train}}
\newcommand{\That}{\hat{\mathcal{T}}}

\newcommand{\SE}{\mathrm{SE}}

\newcommand{\that}{{\hat{t}}}

\newcommand{\M}{\bm{M}}

\renewcommand{\L}{\bm{L}}
\renewcommand{\S}{\bm{S}}
\newcommand{\X}{\bm{S}}
\newcommand{\Y}{\bm{M}}
\newcommand{\W}{\bm{W}}

\newcommand{\tL}{\tilde{\bm{L}}}
\newcommand{\tS}{\tilde{\bm{S}}}

\newcommand{\outfracrow}{\small\text{max-outlier-frac-row}}
\newcommand{\outfraccol}{\small\text{max-outlier-frac-col}}

\newcommand{\SVD}{{SVD}}

\renewcommand{\Re}{\mathbb{R}}

\newcommand{\rmat}{r_L} 

\newcommand{\xmint}{\rho_t}
\newcommand{\init}{\mathrm{init}}

%% file: proc_intro_figs.tex
\pgfplotstableread[col sep = comma]{small_noise_twod.dat}\sninp
\pgfplotstableread[col sep = comma]{large_noise_twod.dat}\lninp

\pgfplotsset{every axis title/.style={text width=.5\textwidth, align=center, at={(.5, -0.2)}, below}}

\begin{figure*}[t!]
\begin{subfigure}[t]{0.5\linewidth}
\centering
\begin{tabular}{@{}ccc@{}}
\\    \newline
\includegraphics[scale=.31]{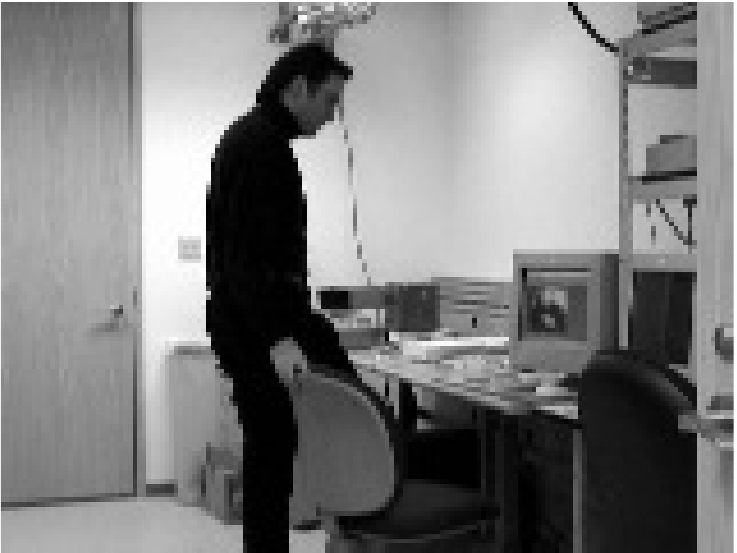} &
\includegraphics[scale=.31]{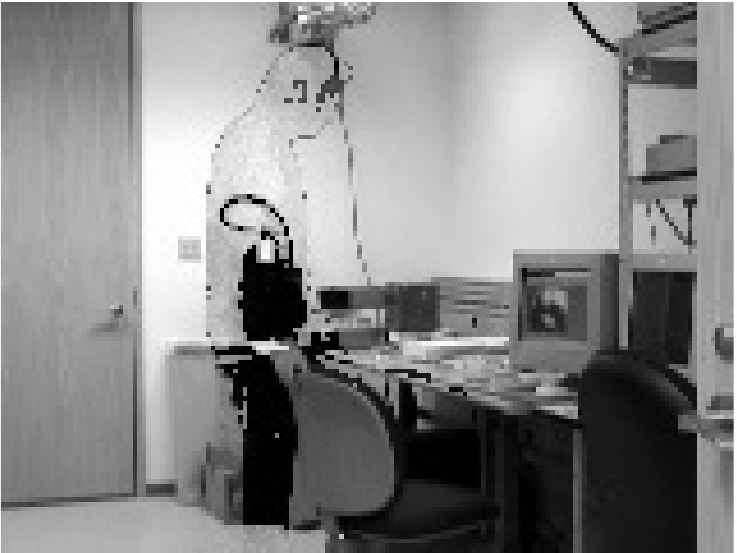} &
\includegraphics[scale=.32]{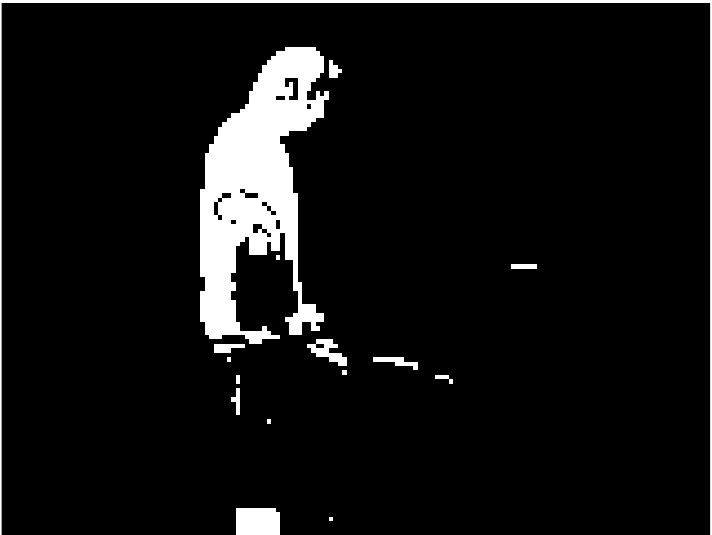} \\ \newline
\includegraphics[scale=.31]{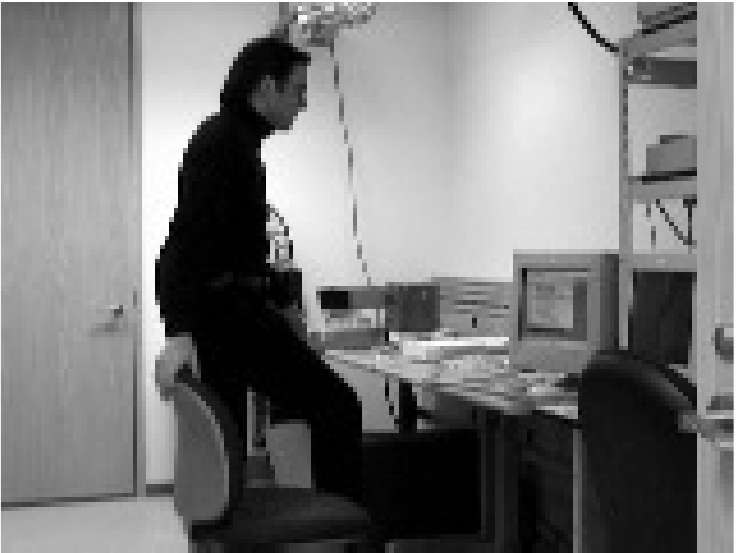} &
\includegraphics[scale=.31]{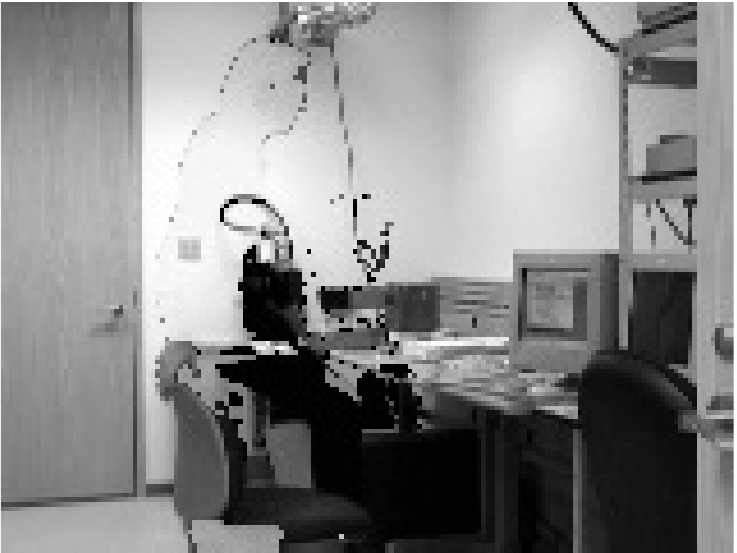} &
\includegraphics[scale=.32]{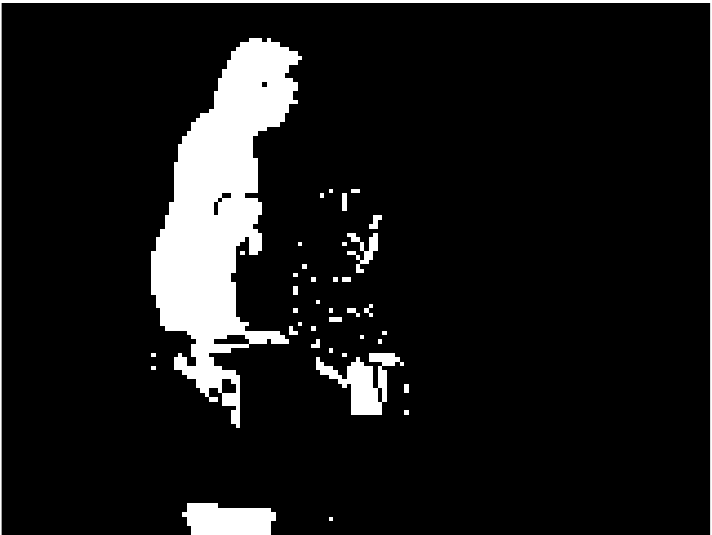} \\ \newline
\includegraphics[scale=.31]{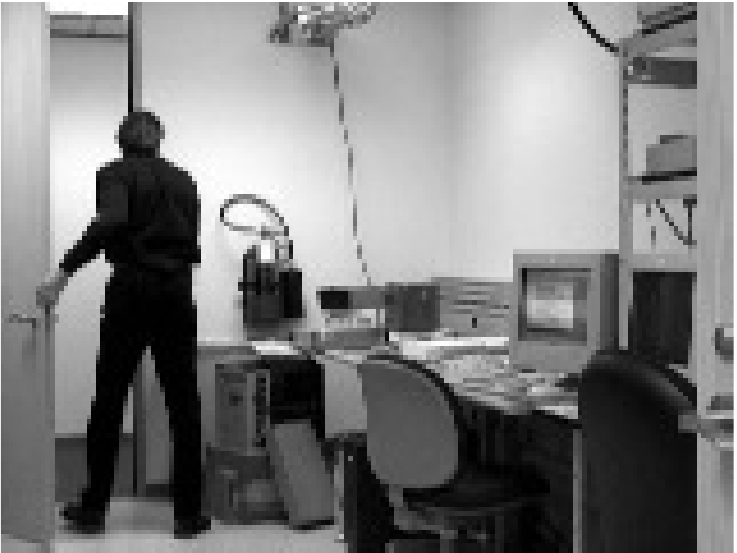} &
\includegraphics[scale=.31]{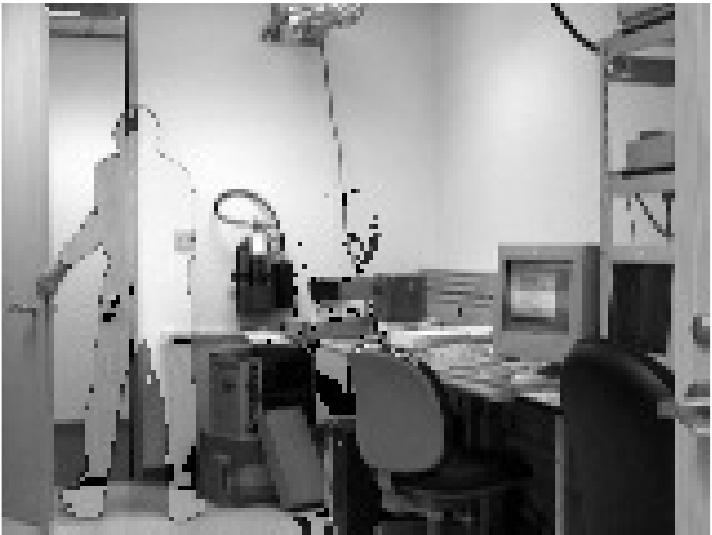} &
\includegraphics[scale=.32]{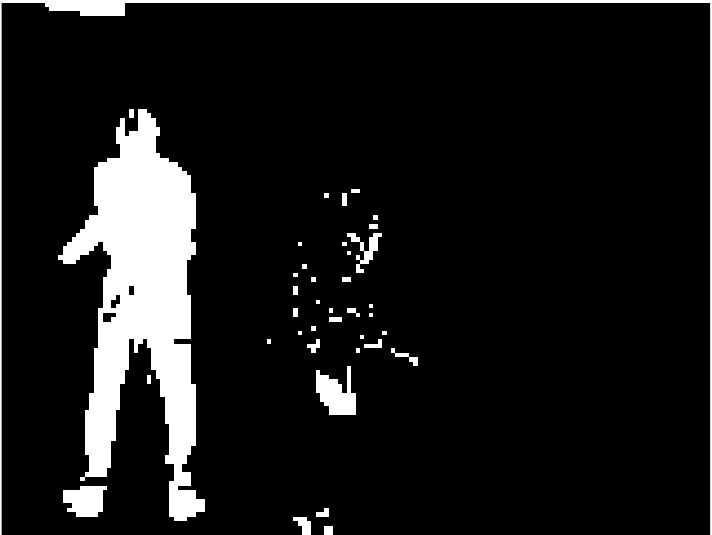}
\\ \newline
\small{Original} & \small{Background} & \small{Foreground}
\end{tabular}
\caption{Video Analytics}
\label{fig_video}
\end{subfigure}%
\begin{subfigure}[t]{0.5\linewidth}
\centering
\begin{tabular}{@{}ccc@{}}
\\    \newline
\includegraphics[scale=0.25]{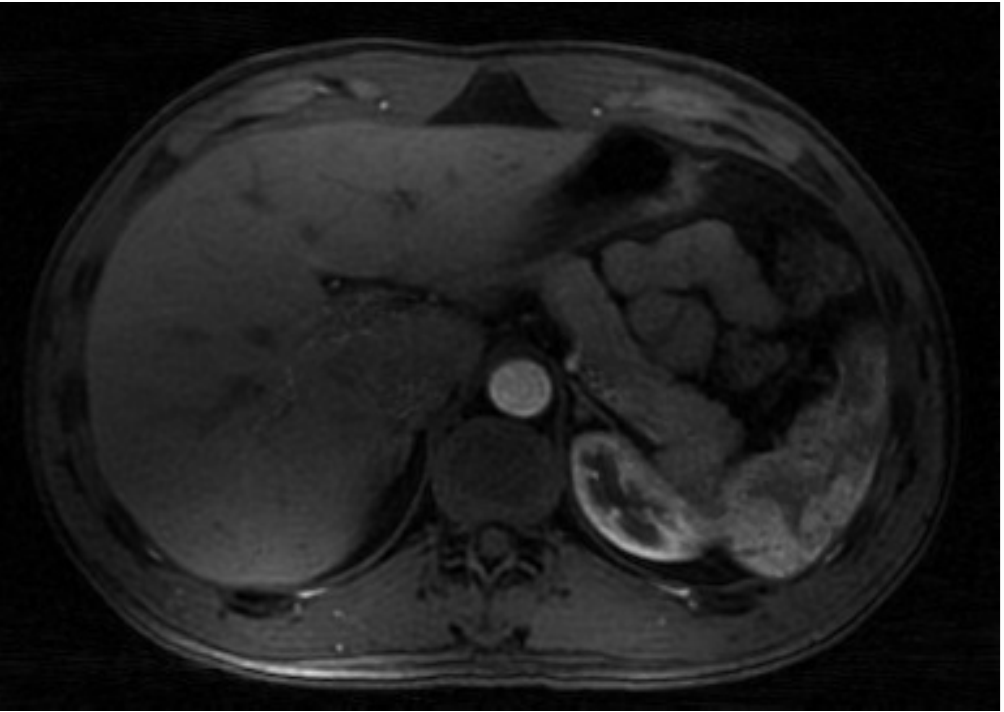} &
\includegraphics[scale=0.25]{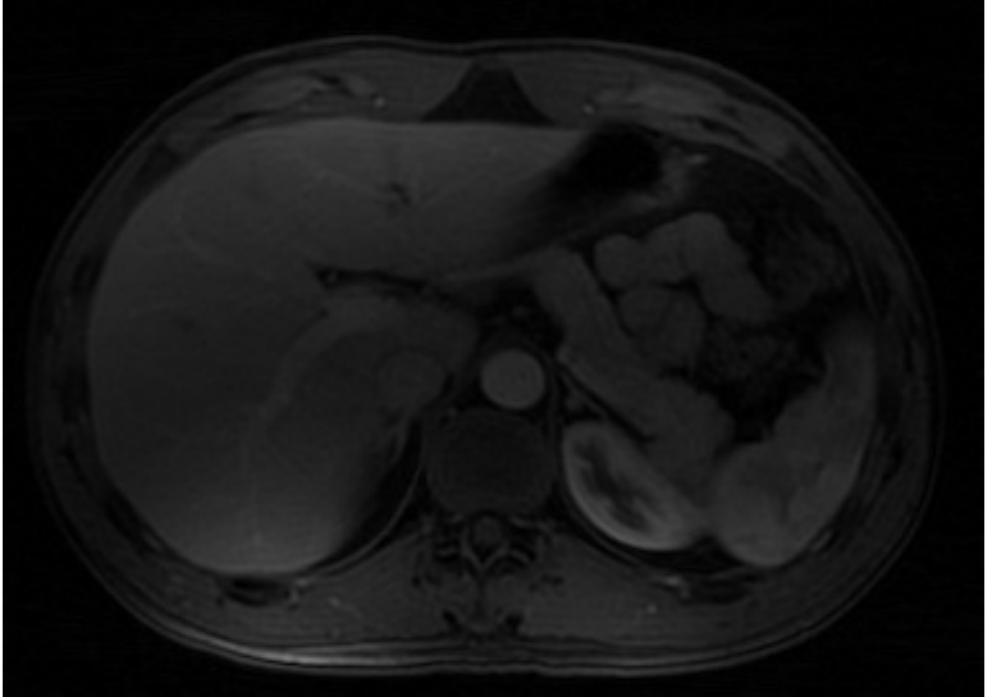} &
\includegraphics[scale=0.25]{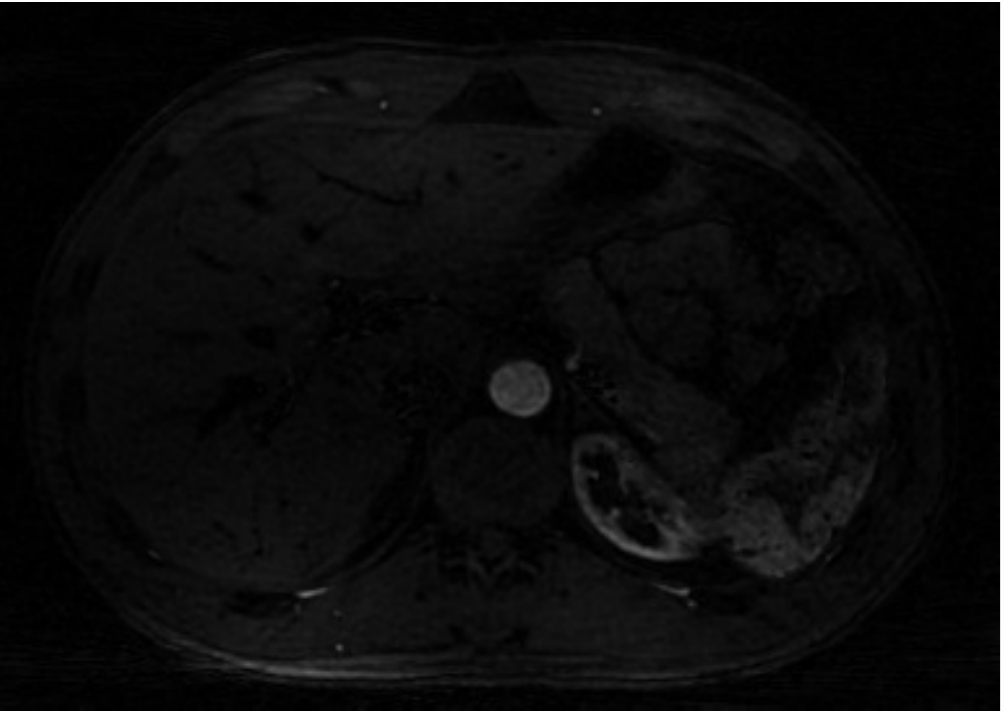} \\ \newline
\includegraphics[scale=0.25]{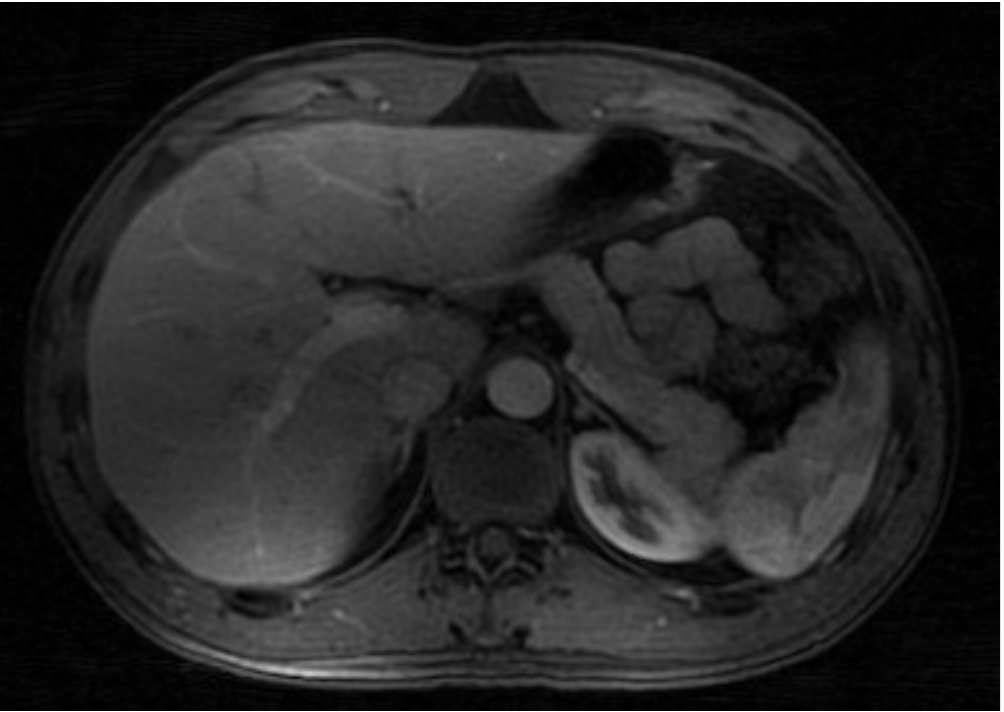} &
\includegraphics[scale=0.25]{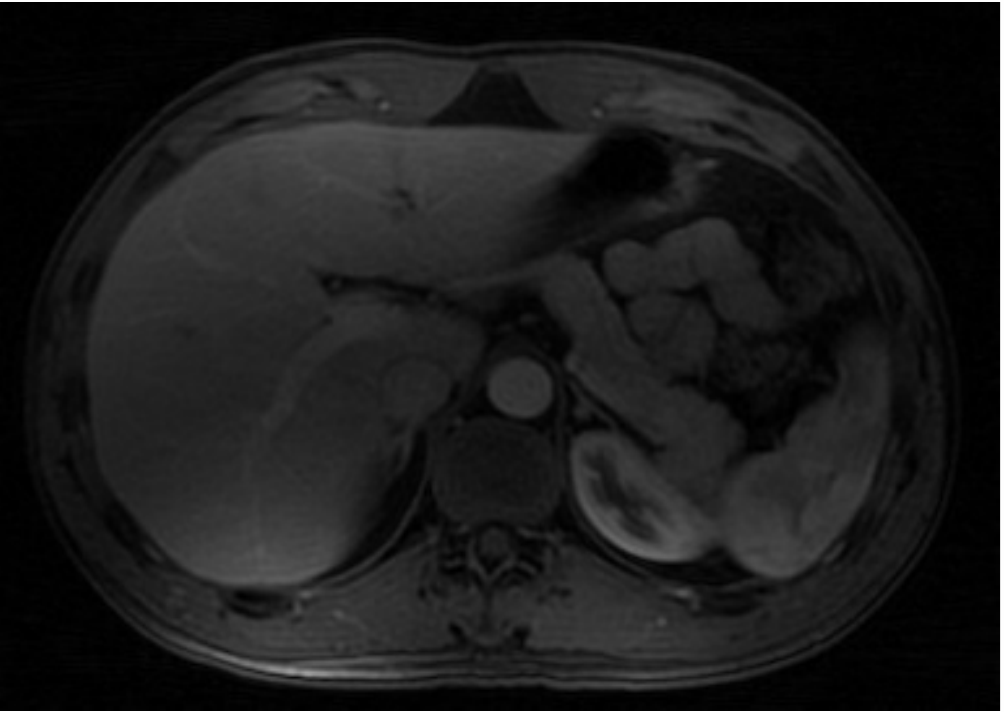} &
\includegraphics[scale=0.25]{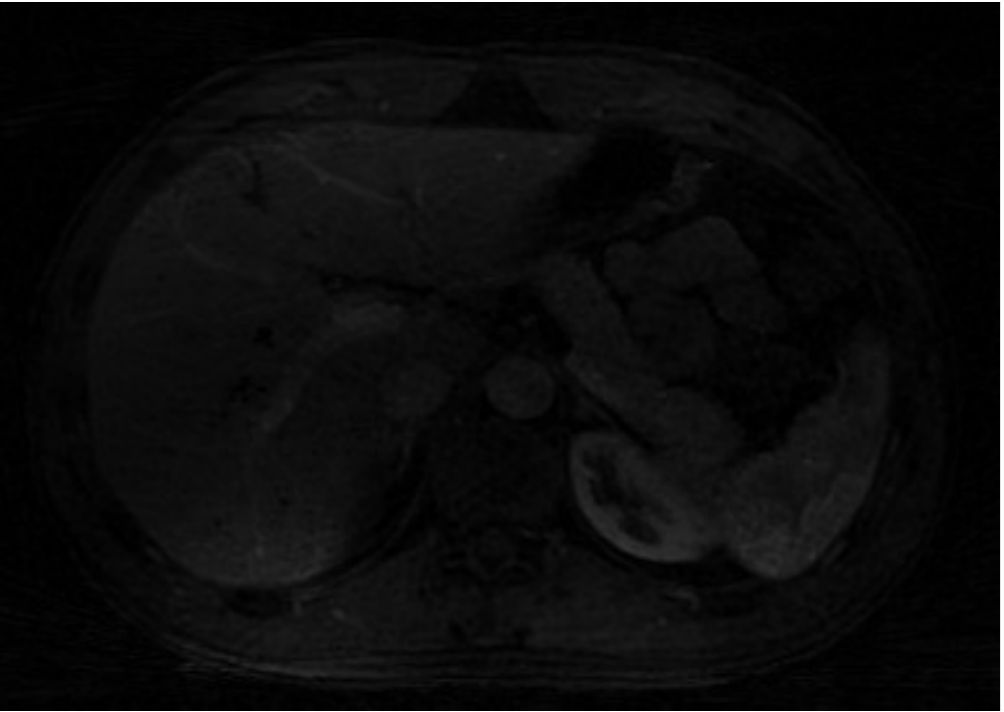} \\ \newline
\includegraphics[scale=0.25]{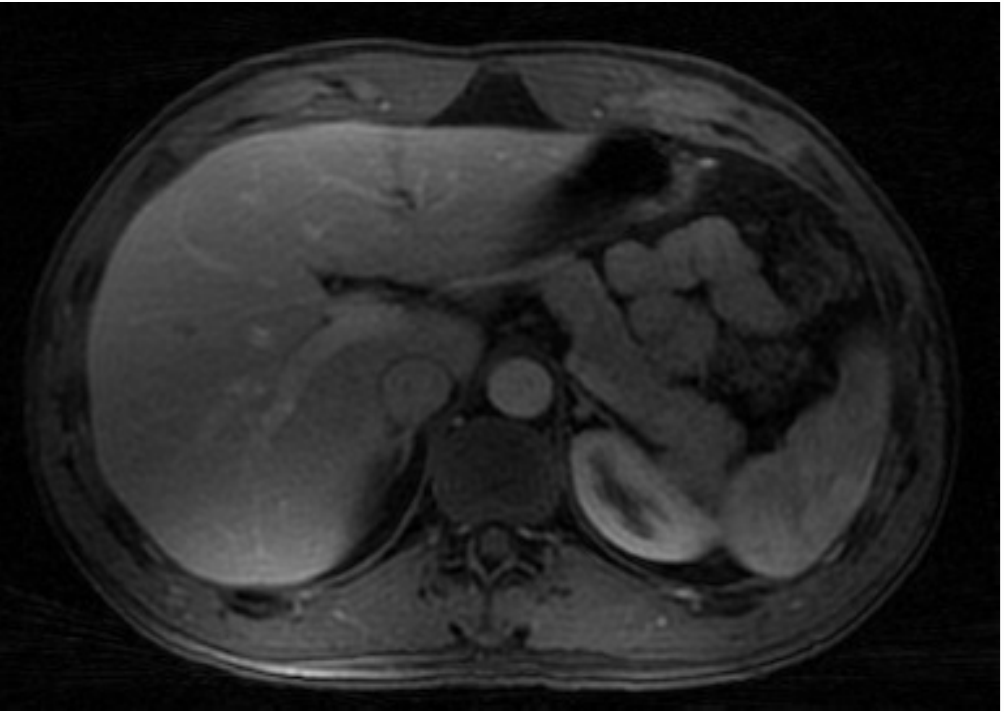} &
\includegraphics[scale=0.25]{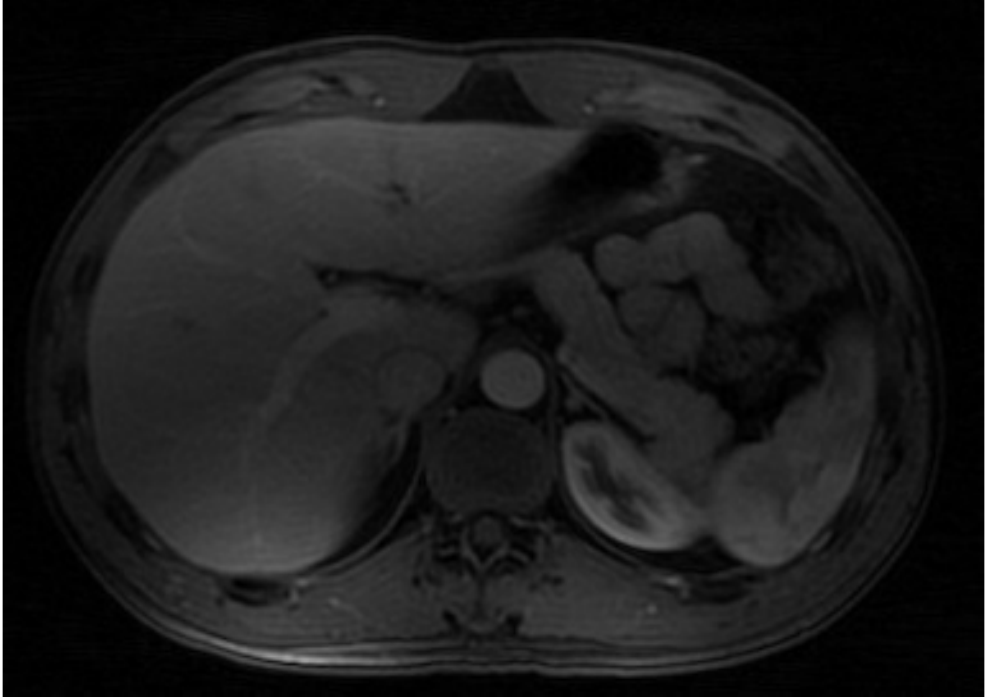} &
\includegraphics[scale=0.25]{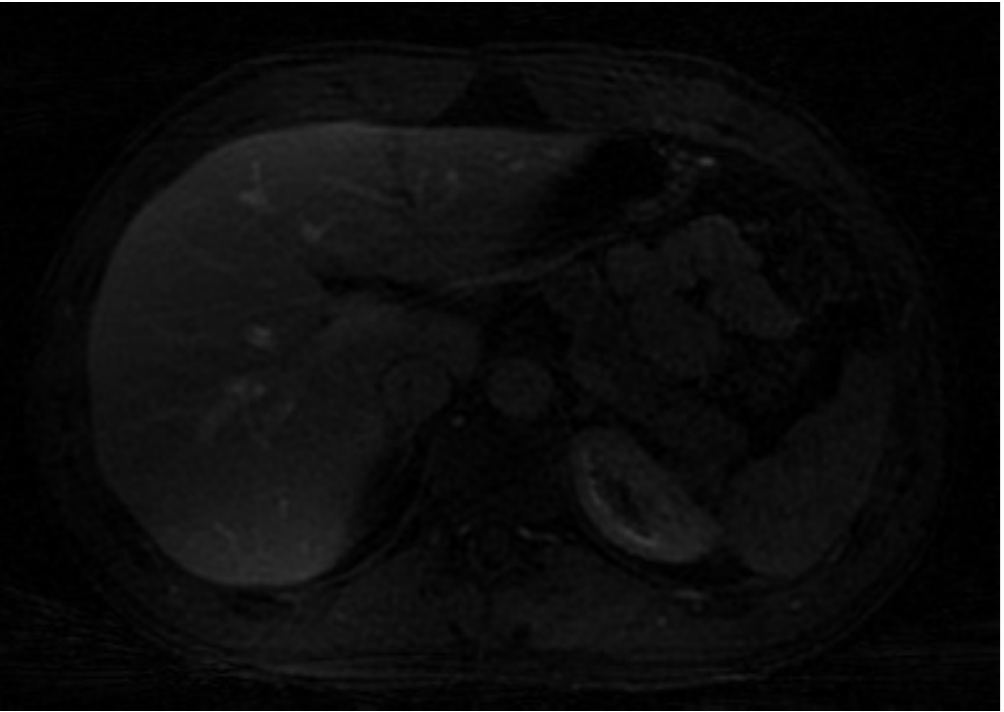} \\ \newline
\small{Original} & \small{Background} & \small{Sparse ROI}
\end{tabular}
\caption{Dynamic MRI}
\label{fig_mri}
\end{subfigure}
\caption{\footnotesize{(a) A video analytics application: Video layering (foreground-background separation) in videos can be posed as a Robust PCA problem. This is often the first step to simplify many computer vision and video analytics' tasks. We show three frames of a video in the first column. The background images for these frames are shown in the second column. Notice that they all look very similar and hence are well modeled as forming a low rank matrix. The foreground support is shown in the third column. This clearly indicates that the foreground is sparse and changes faster than the background. Result taken from \cite{rrpcp_medrop}, code at \url{https://github.com/praneethmurthy/NORST}.
(b) Low-rank and sparse matrix decomposition for accelerated dynamic MRI \cite{candes_mri}. 
The first column shows three frames of abdomen cine data. The second column shows the slow changing background part of this sequence, while the third column shows the fast changing sparse region of interest (ROI). This is also called the ``dynamic component''. These are the reconstructed columns obtained from 8-fold undersampled data. They were reconstructed using under-sampled stable PCP \cite{candes_mri}.
}}
\end{figure*}

%% file: tables_proc_ieee.tex
\begin{table*}[ht!]
\caption{\small{Comparing assumptions, time and memory complexity for static and dynamic RPCA solutions. For simplicity, we ignore all dependence on condition numbers. 
}}
\vspace{-0.12in}
\begin{center}
\renewcommand*{\arraystretch}{1.15}
\resizebox{\linewidth}{!}{
\begin{tabular}{lllll}
\toprule
Algorithm & Outlier tolerance & Assumptions  & Memory, Time, &  \# params.\\
\midrule
PCP(C)\cite{rpca}  & $\outfracrow = \mathcal{O}(1)$& uniform random support & Mem: $\mathcal{O}(n \tmax)$   & zero \\
(offline)        & $\outfraccol = \mathcal{O}(1) $ &       &             Time: $\mathcal{O}(n \tmax^2 \frac{1}{\epsilon})$   & \ \\
        & $\rmat \leq  {c\min(n,\tmax)}/{\log^2 n}$    &     & \   & \ \\
\midrule
AltProj\cite{robpca_nonconvex},  & $\outfracrow = O\left(1/\rmat\right)$ & & Mem: $\mathcal{O}(n \tmax)$   & 2 \\
(offline)   & $\outfraccol = O\left( 1/\rmat\right)$      &    & Time: $\mathcal{O}(n \tmax \rmat^2 \log \frac{1}{\epsilon})$   &   \\
\midrule
RPCA-GD & $\outfracrow = \mathcal{O}(1/\rmat^{1.5})$       &  & Mem: $\mathcal{O}(n \tmax)$   & 5 \\
 \cite{rpca_gd} (offline)       & $\outfraccol = \mathcal{O}(1/\rmat^{1.5})$ & & Time: $\mathcal{O}(n \tmax \rmat \log \frac{1}{\epsilon})$   &  \\
\midrule
NO-RMC& $\outfracrow = O\left(1/\rmat\right)$ & $\tmax  \approx n$  & Mem: $\mathcal{O}(n \tmax)$   & 3 \\
  \cite{rmc_gd} (offline)      & $\outfraccol  = \mathcal{O}(1/\rmat)$            &  & Time: $\mathcal{O}(n \rmat^3 \log^2 n \log^2 \frac{1}{\epsilon})$   &  \\
\midrule
{ReProCS-NORST} & {$\outfracrow  = \mathcal{O}(1) $} & outlier mag. lower bounded       & { Mem: $\mathcal{O}(n \rmat \log n \log \frac{1}{\epsilon} )$ }  & 4 \\
 \cite{rrpcp_merop,rrpcp_medrop} ({\em online})    & $\outfraccol = \mathcal{O}(1/\rmat) $ & slow subspace change or fixed subspace &  Time: $\mathcal{O}(n \tmax \rmat \log \frac{1}{\epsilon})$   & \ \\
 {detects \& tracks}      &   & first $Cr$ samples: AltProj assumptions    &  Detect delay: $ C r \log n$  & \\
{subspace change}   & & \ & & \\
{with near-optimal delay}   & & \ & & \\
%
\hline
%
%
%
%
{simple-ReProCS \cite{rrpcp_dynrpca}:} &        {$\outfracrow  \in \mathcal{O}(1) $} & {subspace change: only 1 direc at a time},      & {Memory: $\mathcal{O}(nr \log n \log \frac{1}{\epsilon} )$}   \\
$r$-times sub-optimal &       {$\outfraccol \in \mathcal{O}(1/r) $} & all ReProCS-NORST assumptions     & {Time: $\mathcal{O}(n \tmax r \log \frac{1}{\epsilon})$}    \\
detect/tracking delay &  & \            &  Detect delay: $C r \log n$   \\
\hline
original-ReProCS \cite{rrpcp_isit15,rrpcp_aistats} &         {$\outfracrow  \in \mathcal{O}(1) $} &  {slow subspace change (unrealistic)}      & {Memory: $\mathcal{O}(nr^2/{\epsilon^2} )$}   \\
$r^2/\epsilon^2$-times sub-optimal &         {$\outfraccol \in \mathcal{O}(1/\rmat) $} & {many unrealistic assumptions}    &  {Time: $\mathcal{O}(n \tmax r \log \frac{1}{\epsilon})$ }   \\
detect/tracking delay &         \ &  all ReProCS-NORST assumptions   & Detect delay: $C n r^2 / \epsilon^2$   \\
\hline
Modified-PCP \cite{zhan_pcp_jp} & $\outfracrow \in \mathcal{O}(1)$ & {outlier support: unif. random} & Memory: $\mathcal{O}(nr \log^2 n)$ \\
piecewise batch                                        & $\outfraccol \in \mathcal{O}(1)$ & {slow subspace change (unrealistic)} & Time: $\mathcal{O}(n \tmax r \log^2 n \frac{1}{\epsilon} )$      \\
tracking soln. & $\rmat \leq {c\min(n,\tmax)}/{\log^2 n}$ &  & Detect delay: $\infty$ \\
\hline
%
\bottomrule
\end{tabular}
}
\label{compare_assu}
\end{center}
\end{table*}

%% file: proc_ieee_expts_3.tex
\newcommand{\lthres}{\omega_{\mathrm{evals}}}

\begin{table*}[t!]
\begin{center}
\caption{\footnotesize{Comparison of $\|\Lhat - \L\|_F/\|\L\|_F$ for Online and offline RPCA methods. Average time for the Moving Object model is given in parentheses. The offline (batch) methods are performed once on the complete dataset.}}
\resizebox{0.9\linewidth}{!}{
\begin{tabular}{cccccccc} \toprule
Outlier Model & GRASTA & ORPCA & {ReProCS-NORST} & RPCA-GD & AltProj & {Offline-NORST} \\  
  & ($0.02$ ms) & ($1.2$ms) & (${0.9}$ {ms}) & ($7.8$ms) & ($4.6$ms) & (${1.7}${ms}) \\  \toprule

ORPCA Model \\ (fixed subspace) & $14.2320$ & $0.0884$ & $0.0921$ & $1.00$ & $0.4246$ & -- \\ \midrule

Moving Object & $0.630$ & $0.861$ & ${4.23 \times 10^{-4}}$ & $4.283$ & $4.441$ & ${8.2 \times 10^{-6}}$ \\
Bernoulli & $6.163$ & $1.072$ & ${0.002}$ & $0.092$& $0.072$ & ${2.3 \times 10^{-4}}$ \\ \bottomrule
\end{tabular}
}
\label{tab:offline}
\end{center}
\end{table*}


\pgfplotstableread[col sep = comma]{figures/final_files/MO_online_finalSE_TIT.dat}\ltmodata
\pgfplotstableread[col sep = comma]{figures/final_files/bern_online_finalSE_TIT.dat}\ltmodatabern
\pgfplotstableread[col sep = comma]{figures/final_files/norst_fail_all_bigalpha.dat}\nfdata
\pgfplotstableread[col sep = comma]{figures/final_files/norst_fail_change.dat}\nfcdata

\begin{figure*}[t!]
\centering
\begin{tikzpicture}
    \begin{groupplot}[
        group style={
            group size=2 by 2,
            y descriptions at=edge left,
            vertical sep=2cm
        },
        my stylecompare,
        width = .48\linewidth,
        height=4cm,
    ]
           \nextgroupplot[
            legend entries={
            	GRASTA - $0$,
            	GRASTA - $100$,
            	GRASTA - $200$,
            	AltProj,
            	GradDesc,
            	ORPCA,
            	ReProCS-NORST - $100$,
            	ReProCS-NORST - $200$,
            		},
            legend style={at={(2.15,1.45)}},
            legend columns = 8,
            legend style={font=\tiny}, 
            xlabel=$t$,
            ymode=log,
            ylabel={\small{$\SE(\Phat_{(t)}, \P_{(t)})$}},
            title={(a)},
            title style={at={(0.3, -.3)}},
            ymin=13e-3, ymax = 1,
        ]
			\addplot[red, line width=1.6pt, mark=square,mark size=3.5pt] table[x index = {0}, y index = {1}]{\nfdata};
   	        \addplot[red, line width=1.6pt, mark=o,mark size=3.5pt] table[x index = {0}, y index = {2}]{\nfdata};
   	        \addplot[red, line width=1.6pt, mark=Mercedes star,mark size=3.5pt] table[x index = {0}, y index = {3}]{\nfdata};
	        \addplot[cyan, line width=1.6pt, mark=o,mark size=3.5pt] table[x index = {0}, y index = {4}]{\nfdata};
	        \addplot[olive, line width=1.6pt, mark=square,mark size=3pt] table[x index = {0}, y index = {5}]{\nfdata};
	        \addplot[teal, line width=1.6pt, mark=o,mark size=3.5pt] table[x index = {0}, y index = {6}]{\nfdata};
	        \addplot[blue, line width=1.6pt, mark=square,mark size=3pt] table[x index = {0}, y index = {7}]{\nfdata};
	        \addplot[blue, line width=1.6pt, mark=o,mark size=3pt] table[x index = {0}, y index = {8}]{\nfdata};
	
           \nextgroupplot[
            xlabel=$t$,
            title={(b)},
            title style={at={(0.3, -.3)}},
            ymin=13e-3, ymax = 1,
            ymode=log
        ]
			\addplot[red, line width=1.6pt, mark=square,mark size=3.5pt, mark repeat=2] table[x index = {0}, y index = {1}]{\nfcdata};
   	        \addplot[teal, line width=1.6pt, mark=o,mark size=3.5pt, mark repeat=2] table[x index = {0}, y index = {2}]{\nfcdata};
   	        \addplot[blue, line width=1.6pt, mark=o,mark size=3.5pt, mark repeat=2] table[x index = {0}, y index = {3}]{\nfcdata};
	        \addplot[cyan, line width=1.6pt, mark=o,mark size=3.5pt, mark repeat=2] table[x index = {0}, y index = {4}]{\nfcdata};
	        \addplot[olive, line width=1.6pt, mark=square,mark size=3pt, mark repeat=2] table[x index = {0}, y index = {5}]{\nfcdata};

           \nextgroupplot[
            my legend style compare,
            legend style={at={(.4,1.3)}},
            legend columns = 7,
        enlargelimits=false,
            xlabel=$t$,
            ylabel={\small{$\log \SE(\Phat_{(t)}, \P_{(t)})$}},
            title={(c)},
            title style={at={(0.3, -.3)}},
        ]
	        \addplot table[x index = {0}, y index = {1}]{\ltmodata};
	        \addplot table[x index = {0}, y index = {2}]{\ltmodata};
	        \addplot table[x index = {0}, y index = {4}]{\ltmodata};
	        \addplot table[x index = {0}, y index = {5}]{\ltmodata};
		               \nextgroupplot[
            ymode=log,
            xlabel={\small{$t$}},
            title={(d)},
            title style={at={(0.3, -.3)}},
            enlargelimits=false,
        ]
	        \addplot table[x index = {0}, y index = {1}]{\ltmodatabern};
	        \addplot table[x index = {0}, y index = {2}]{\ltmodatabern};
	        \addplot table[x index = {0}, y index = {4}]{\ltmodatabern};
	        \addplot table[x index = {0}, y index = {5}]{\ltmodatabern};

    \end{groupplot}
\end{tikzpicture}
\caption{\footnotesize{Figure illustrating the results on synthetic data. \textbf{(a)}: Plot of subspace error in the case of a fixed subspace. GRASTA (ReProCS-NORST) - $100$, $200$ indicate that the initialization used $t_\train = 100, 200$ samples. GRASTA-$0$ indicates that we used the default initialization (to zeros). Notice that ORPCA and ReProCS-NORST-200 are able to improve the subspace estimate using more time samples, while others fail. ReProCS-NORST-100 failed because the initial subspace error was one (too large). All versions of GRASTA fail too. {\bf (b)}: Illustrates the subspace error for the ORPCA-model, but with changing subspace. ORPCA and ReProCS-NORST are able to obtain good accuracy. {\bf (c)}: Illustrates the subspace error for outlier supports generated using Moving Object Model and {\bf (d)}: illustrates the error under the Bernoulli model. The values are plotted every $k\alpha - 1$ time-frames. The time taken per frame in milliseconds (ms) for the Bernoulli model is shown in legend parentheses. All results are averaged over $100$ independent trials. 
{\em The key difference between the first two plots and the last two plots is that (i) in the first two plots, the initial error seen by ReProCS-NORST-200 is much higher (around 0.98); and (ii) outliers are generated to be uniformly distributed between -1000 and 1000 (and so many outliers are neither too large, nor too small). 
}
}}
\vspace{-0.1in}
\label{fig:norst_fail}
\end{figure*}
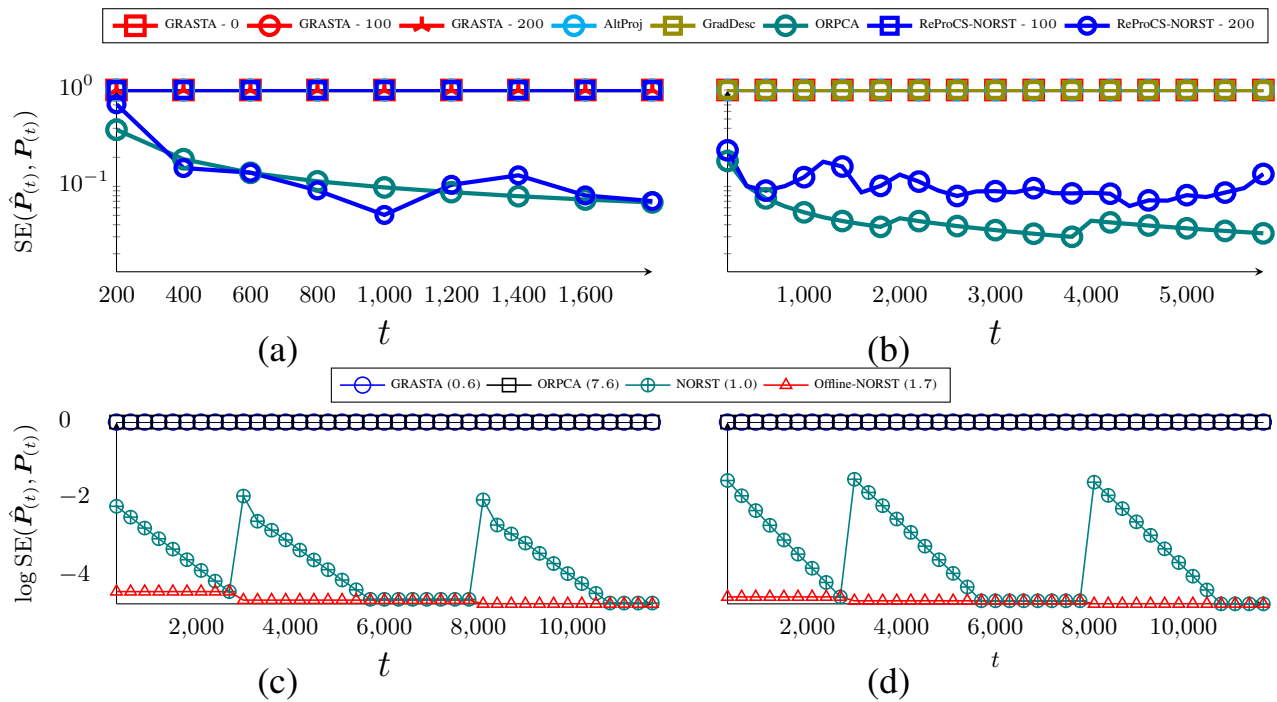

\section{Experimental comparisons}\label{sims}
All time comparisons are performed on a Desktop Computer with Intel$^{\textsuperscript{\textregistered}}$ Xeon E$3$-$1240$ $8$-core CPU @ $3.50$GHz and $32$GB RAM. All experiments on synthetic data are averaged over $100$ independent trials.

\subsubsection{Synthetic Data: Fixed Subspace}
Our first experiment generates data exactly as suggested in the ORPCA paper \cite{xu_nips2013_1}. The true data subspace is {\em fixed}. We generate the low rank matrix $\bm{L} = \bm{U} \bm{V}{}'$ where the entries of $\bm{U} \in \mathbb{R}^{n \times r}$ and $\bm{V}\in \mathbb{R}^{\tmax \times r}$ are generated as i.i.d $\mathcal{N}(0, 1/\tmax)$. Notice that in this case, $\P_{(t)} := \basis(\bm{U})$.
We used the Bernoulli model to generate the sparse outliers. This means that each entry of the $n \times \tmax$  sparse outlier matrix $\X$ is nonzero with probability $\rho_x$ independent of all others. The nonzero entries  are generated uniformly at random in the interval $[-1000, 1000]$. We used $n=400$, $\tmax = 1000$, $r = 50$, and $\rho_x = 0.001$. The setting stated in \cite{xu_nips2013_1} used $\rho_x = 0.01$, but we noticed that in this case, ORPCA error saturated at $0.4$. To show a case where ORPCA works well, we reduced $\rho_x$ to 0.001. We compare ReProCS, ORPCA and GRASTA. ReProCS-NORST given earlier in Algorithm \ref{norst_basic} was implemented. 

ORPCA does not use any initialization, while ReProCS does. GRASTA has the option of providing an initial estimate or using the default initialization of zero. In this first experiment we tried both options.
The initial subspace estimate for both ReProCS and GRASTA was computed using AltProj applied to the first $t_\train$ frames. We experimented with two values of $t_\train$: $t_\train=100$ and $t_\train=200$. We label the corresponding algorithms ReProCS-100, ReProCS-200, GRASTA-100, and GRASTA-200. GRASTA with no initialization provided is labeled as GRASTA-0.
The Monte Carlo averaged subspace recovery error $\SE(\Phat_{(t)},\P_{(t)})$ versus time $t$ plots are shown in Fig. \ref{fig:norst_fail}. As can be seen, ORPCA works well in this setting while all three versions of GRASTA fail completely.
ReProCS-100 also fails. This is because when $t_\train = 100$, the initial subspace estimate computed using AltProj satisfies $\SE(\Phat_\init, \P_0) = 1$. On the other hand, ReProCS-200 works as well as ORPCA because in this case, $\SE(\Phat_\init, \P_0) \approx 0.98$.

We implement GRASTA and ORPCA using code downloaded from \url{https://github.com/andrewssobral/lrslibrary}. The regularization parameter for ORPCA was set as with $\lambda_1 = 1 / \sqrt{n}$ and $\lambda_2 = 1 / \sqrt{d}$ according to \cite{xu_nips2013_1}.
The ReProCS algorithm parameters are set as suggested in Algorithm \ref{norst_basic}: $K = \lceil \log(c/\varepsilon) \rceil = 8$, $\alpha = C r \log n = 200$, $\omega = 1$ and $\xi = 2/15 = 0.13$, $\lthres = 2 \varepsilon^2 \lambda^+ = 0.00075$.
%

\begin{figure*}[t!]
\begin{center}
\resizebox{\linewidth}{!}{
\begin{tabular}{@{}c@{}c@{}c@{}c@{}c@{}c@{}}
\\    \newline
	\includegraphics[width=0.11\linewidth, height=1.7cm]{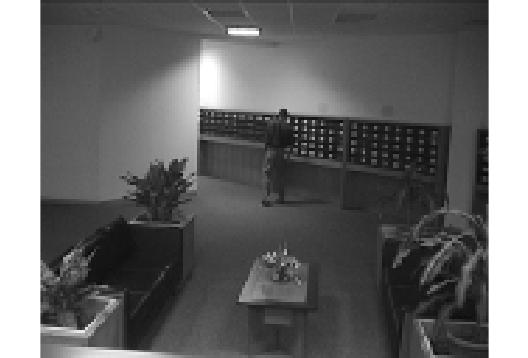}
&
	\includegraphics[width=0.11\linewidth, height=1.7cm, trim={.7cm, 0cm, .7cm, 0cm}, clip]{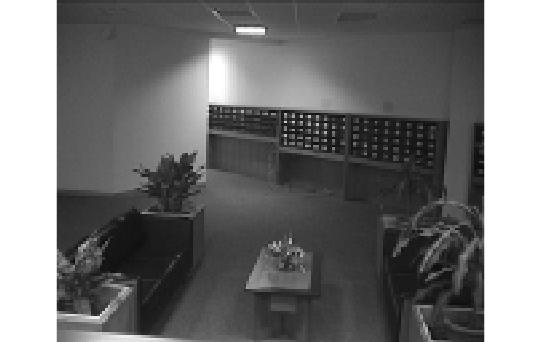}
&
	\includegraphics[width=0.11\linewidth, height=1.7cm, trim={.7cm, 0cm, .7cm, 0cm}, clip]{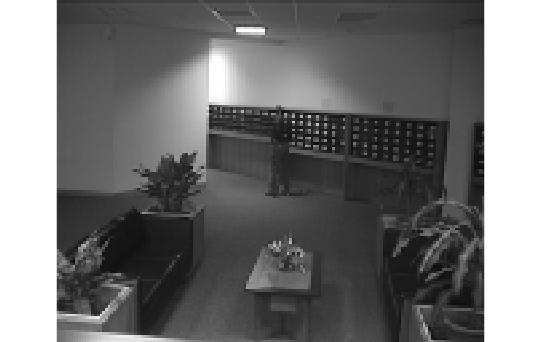}
&
	\includegraphics[width=0.11\linewidth, height=1.7cm, trim={.7cm, 0cm, .7cm, 0cm}, clip]{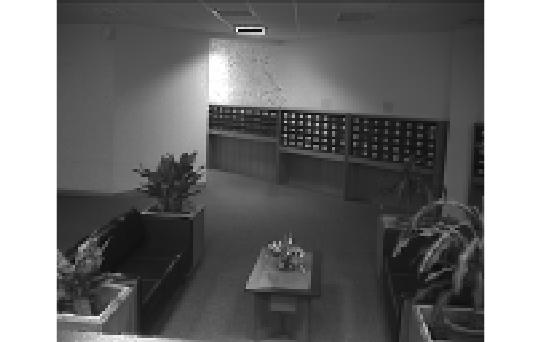}
&
	\includegraphics[width=0.11\linewidth, height=1.7cm, trim={.7cm, 0cm, .7cm, 0cm}, clip]{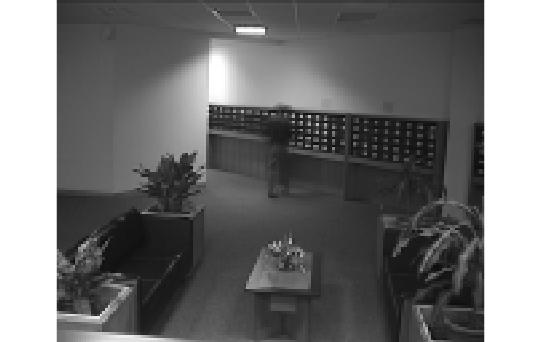}
&
	\includegraphics[width=0.11\linewidth, height=1.7cm, trim={1.25cm, 0cm, 1.2cm, 0cm}, clip]{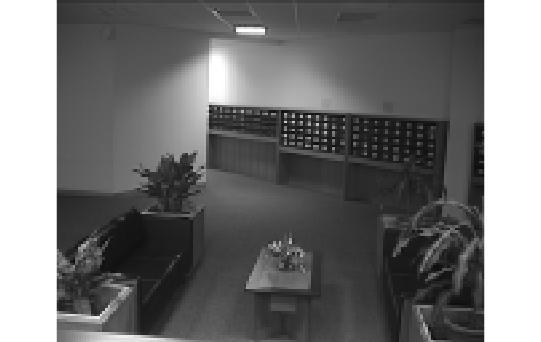}
\\    \newline
	{\includegraphics[width=0.11\linewidth, height=1.7cm, trim={.8cm, 0cm, 1cm, 0cm}, clip]{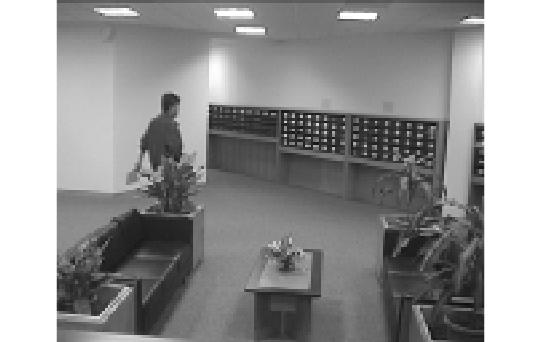}}
&
	{\includegraphics[width=0.11\linewidth, height=1.7cm, trim={.7cm, 0cm, .7cm, 0cm}, clip]{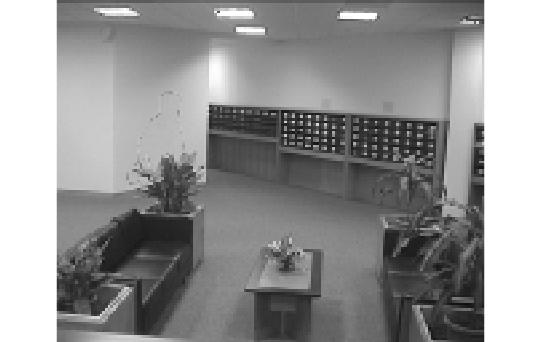}}
&
	{\includegraphics[width=0.11\linewidth, height=1.7cm, trim={1.25cm, 0cm, 1.2cm, 0cm}, clip]{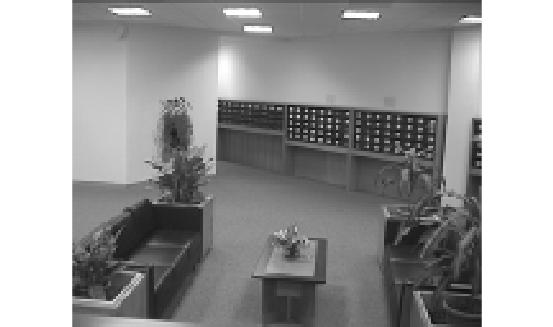}}
&
	{\includegraphics[width=0.11\linewidth, height=1.7cm, trim={1.2cm, 0cm, 1.2cm, 0cm}, clip]{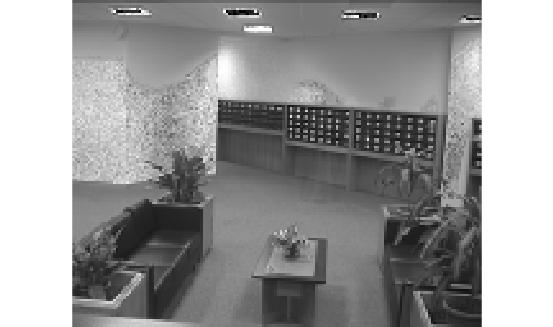}}
&
	{\includegraphics[width=0.11\linewidth, height=1.7cm, trim={1.2cm, 0cm, 1.25cm, 0cm}, clip]{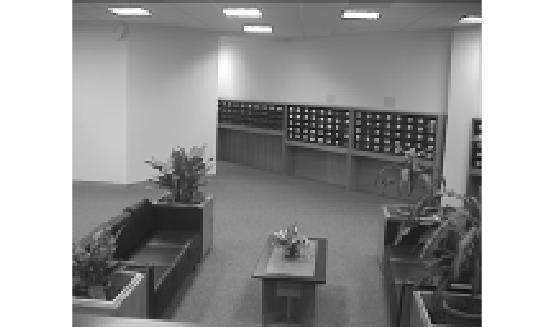}}
&
	{\includegraphics[width=0.11\linewidth, height=1.7cm, trim={.7cm, 0cm, .7cm, 0cm}, clip]{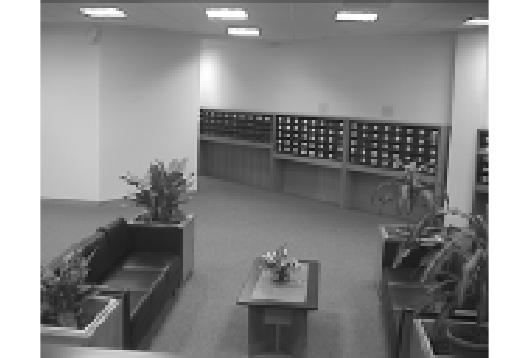}}
		\\ \newline 
			\tiny{Original} & \tiny{ReProCS($16.5$ms)} & \tiny{AltProj($26.0$ms)} & \tiny{RPCA-GD($29.5$ms)} & \tiny{GRASTA($2.5$ms)} & \tiny{PCP ($44.6$ms)} \\ \newline		
		\includegraphics[width=0.11\linewidth, height=1.7cm, trim={1.2cm, 0cm, 1.25cm, 0cm}, clip]{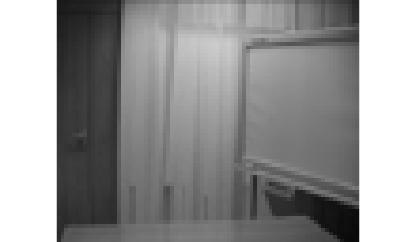}
&
	\includegraphics[width=0.11\linewidth, height=1.7cm, trim={1.2cm, 0cm, 1.25cm, 0cm}, clip]{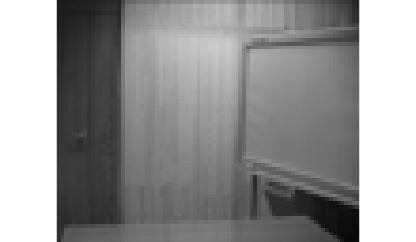}
&
	\includegraphics[width=0.11\linewidth, height=1.7cm, trim={1.1cm, 0cm, 1.15cm, 0cm}, clip]{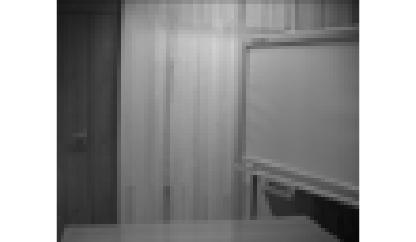}
&
	\includegraphics[width=0.11\linewidth, height=1.7cm, trim={1.05cm, 0cm, 1.25cm, 0cm}, clip]{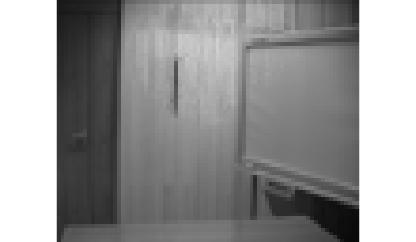}
&
	\includegraphics[width=0.11\linewidth, height=1.7cm, trim={0.38cm, 0cm, 0.4cm, 0cm}, clip]{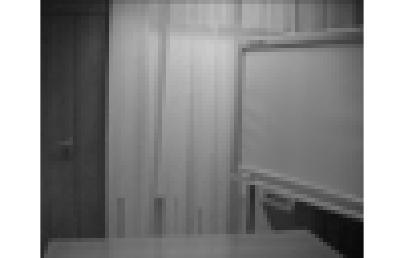}
&
	\includegraphics[width=0.11\linewidth, height=1.7cm, trim={2.1cm, 0cm, 2.2cm, 0cm}, clip]{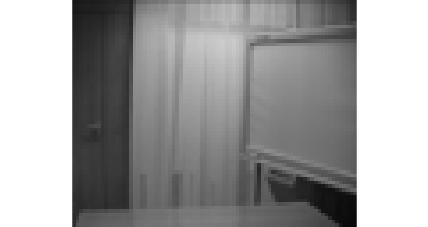}
\\    \newline
	{\includegraphics[width=0.11\linewidth, height=1.7cm, trim={1.8cm, 0cm, 1.8cm, 0cm}, clip]{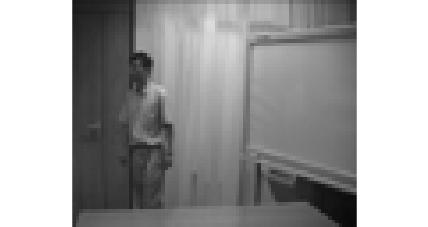}}
&
	{\includegraphics[width=0.11\linewidth, height=1.7cm, trim={1.2cm, 0cm, 1.25cm, 0cm}, clip]{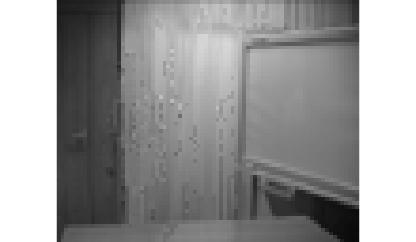}}
&
	{\includegraphics[width=0.11\linewidth, height=1.7cm, trim={1.7cm, 0cm, 1.7cm, 0cm}, clip]{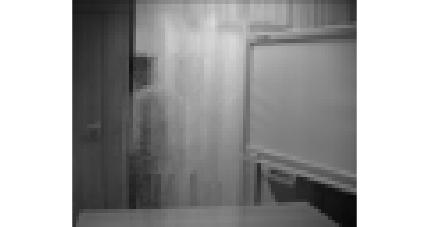}}
&
	{\includegraphics[width=0.11\linewidth, height=1.7cm, trim={1.65cm, 0cm, 1.8cm, 0cm}, clip]{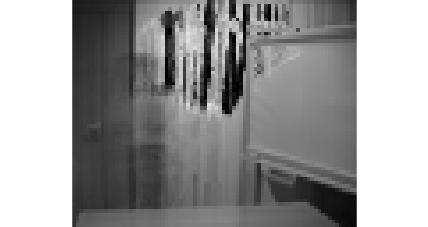}}
&
	{\includegraphics[width=0.11\linewidth, height=1.7cm, trim={1cm, 0cm, 1.15cm, 0cm}, clip]{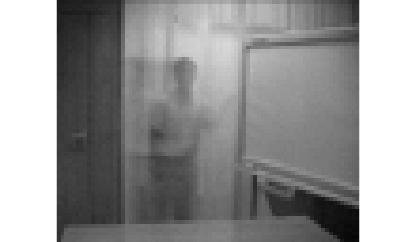}}
&
	{\includegraphics[width=0.11\linewidth, height=1.7cm, trim={1.5cm, 0cm, 1.6cm, 0cm}, clip]{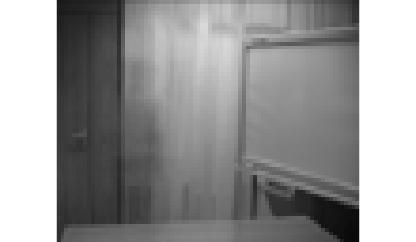}}  \\ \newline
	\tiny{Original} & \tiny{ReProCS($72.5$ms)} & \tiny{AltProj($133.1$ms)} & \tiny{RPCA-GD($113.6$ms)} & \tiny{GRASTA($18.9$ms)}  & \tiny{PCP($240.7$ms)} \\
\end{tabular}
}
\caption{\footnotesize{Comparison of background recovery performance is Foreground-Background Separation tasks for the Lobby video (first two rows) and the Meeting Room video (last two rows). The recovered background images are shown at $t= t_\train + 260, 610$ for LB and $t = t_\train + 10, 140$ for MR. The time taken for each of the videos per frame in milliseconds is given in parenthesis. The lobby video involves a nearly static background and hence all algorithms work. The meeting room video involves much more significant background video changes due to the moving curtains. As can be seen, in this case, GRASTA fails. Also, ReProCS-NORST (abbreviated to ReProCS in this figure) is the second fastest after GRASTA and the fastest among solutions that work well.
}}
\label{realvid}
\end{center}
\end{figure*}

To address a comment of the reviewer, we also tried to generate data using the data generation code for GRASTA. However even in this case, the final subspace recovery error of GRASTA was around $0.8$ even when the number of observed data vectors was $\tmax = 12000$. We even tried changing its multiple tuning parameters, but were unable to find any useful setting. Further, the code documentation does not provide a good guide to tune the internal parameters for optimization in different scenarios and thus the results here are reported as is. This is possibly since the algorithm parameters on the website are tuned for the video application but the parameter set is not a good setting for synthetic data.

\subsubsection{Synthetic Data: Time-Varying Subspaces (a)}
In our second experiment, we assume that the subspace changes every so often and use $t_j$ to denote the $j$-th change time for $j=1,2,\dots,J$ with $t_0:=1$ and $t_{J+1}:=\tmax$. As explained earlier, this is necessary for identifiability of all the subspaces. In the first sub-part, we use the data generation model as described in the static subspace case with the outliers, outlier magnitudes being generated exactly the same way. For the low-rank part, we simulate the changing subspace as follows. For $t \in [1, t_1]$ we generate $\bm{U}_1$ as in the first experiment. For $t \in (t_1, t_2]$, we use $\bm{U}_2 = \exp(\gamma \bm{B}) \bm{U}_1$, where $\gamma = 10^{-3}$ and $\bm{B}$ is a skew-symmetric matrix, and for $t\in (t_2, \tmax]$ we use $\bm{U}_3 = \exp(\gamma \bm{B}) \bm{U}_2$. The matrix $\bm{V}$ is generated exactly the same as in the first experiment. The results are shown in Fig. \ref{fig:norst_fail}. We note here that ORPCA and ReProCS-NORST provide a good performance and ORPCA is the best as the number of samples increases. All algorithms were implemented using the same parameters described in the first experiment. Here we used $n=400$, $\tmax = 6000$, $r=50$ and $\rho_x = 0.001$. From the previous experiment, we selected GRASTA-0 and ReProCS-200 since these provide best performance and the other algorithm parameters are unchanged.

\subsubsection{Synthetic Data: Time-Varying Subspaces(b)}
In our final two experiments, we again assume that the subspace changes every so often and use $t_j$ to denote the $j$-th change time, Thus, in this case, $\l_t = \P_{(t)} \a_t$ where $\P_{(t)}$ is an $n \times r$ basis matrix with $\P_{(t)} = \P_j$ for $t \in [t_j, t_{j+1})$, $j=0,1,2,\dots, J$.
We generated $\P_0$ by orthonormalizing and $n\times r$ i.i.d. Gaussian matrix. For $j>1$, the  basis matrices $\P_j$ were generated using the model as also done in \cite{grass_undersampled} which involves a left-multiplying the basis matrix with a rotation matrix, i.e.,
\begin{align*}
\P_j = e^{\delta_j \bm{B}_j} \P_{j-1}
\end{align*}
where $\bm{B}_j$ is a skew-Hermitian matrix which ensures that $\P_j{}' \P_j = \I_r$ and $\delta_j$ controls the amount of subspace change. The matrices $\bm{B}_{1}$ and $\bm{B}_2$ are generated as $\bm{B}_{1} = (\tilde{\bm{B}}_1 - \tilde{\bm{B}}_1{}')$ and $\bm{B}_2 = (\tilde{\bm{B}}_2 - \tilde{\bm{B}}_2{}')$ where the entries of $\tilde{\bm{B}}_1, \tilde{\bm{B}}_2$ are generated independently from a standard normal distribution.
To obtain the low-rank matrix $\bm{L}$ from this we generate the coefficients $\at \in \mathbb{R}^{r_0}$ as independent zero-mean, bounded random variables. They are $(\at)_i \overset{i.i.d}{\sim} unif[-q_i, q_i]$ where $q_i = \sqrt{f} - \sqrt{f}(i-1)/2r$ for $i = 1, 2, \cdots, r - 1$ and $q_{r} = 1$ thus the condition number is $f$ and we selected $f=50$.
We used the following parameters: $n = 1000$, $d = 12000$, $J = 2$, $t_1 = 3000$, $t_2 = 8000$, $r = 30$, $\delta_1 = 0.001$, $\delta_2 = \delta_1$.

The sparse outlier matrix $\X$ was generated using two models: (A) Bernoulli model (commonly used model in all RPCA works) and (B) the moving object model \cite[Model G.24]{rrpcp_dynrpca}. This model was introduced in \cite{rrpcp_isit15,rrpcp_dynrpca} as one way to generate data with a larger $\outfracrow^{\alpha}$ than $\outfraccol$. It simulates a person pacing back and forth in a room. The nonzero entries of $\X$ were generated as uniformly at random from the interval $[\xmint, x_{\max}]$ with $\xmint = 10$ and $x_{\max} = 20$ (all independent). With both models we generated data to have fewer outliers in the first $t_\train=100$ frames and more later. This was done to ensure that the batch initialization provides a good initial subspace estimate. 
With the Bernoulli model, we used $\rho_x = 0.01$ for the first $t_\train$ frames and $\rho_x = 0.3$ for the subsequent frames. With the moving object model we used $s/n = 0.01$, $b_0 = 0.01$ for the first $t_\train$ frames and $s/n = 0.05$ and $b_0 = 0.3$ for the subsequent frames. The  subspace recovery error plot is shown in Fig. \ref{fig:norst_fail} (b) (Bernoulli outlier support) and (c) (Moving Object outlier support), while the average $\|\Lhat - \L\|_F / \|\L\|_F$  is compared in Table \ref{tab:offline}. In this table, we compare all RPCA and dynamic RPCA solutions (AltProj, RPCA-GD, ReProCS, ORPCA, GRASTA). We do not compare PCP since it is known to be very slow from all past literature \cite{robpca_nonconvex,rrpcp_dynrpca}.

This experiment shows that since the outlier fraction per row is quite large, the other techniques are not able to obtain meaningful estimates. It also shows that both ReProCS and its offline version are the fastest among all methods that work. They are slower than only ORPCA and GRASTA which never work in either of these experiments.

We initialized ReProCS-NORST using AltProj applied to $\Y_{[1,t_\train]}$ with $t_\train=100$. AltProj used the true value of $r$, $10$ iterations and a threshold of $0.01$. This, and the choice of $\delta_1$ and $\delta_2$ ensure that $\SE(\Phat_\init, \P_0) \approx \SE(\P_1, \P_0) \approx \SE(\P_2, \P_1) \approx 0.01$. The other algorithm parameters are set as mentioned in the algorithm i.e., $K = \lceil \log(c/\varepsilon) \rceil = 8$, $\alpha = C r \log n = 300$, $\omega = \xmint/2 = 5$ and $\xi = \xmint/15 = 0.67$, $\lthres = 2 \varepsilon^2 \lambda^+ = 7.5 \times 10^{-4}$. 
We implement the other algorithms using code downloaded from \url{https://github.com/andrewssobral/lrslibrary}. The regularization parameter for ORPCA was set as with $\lambda_1 = 1 / \sqrt{n}$ and $\lambda_2 = 1 / \sqrt{d}$ according to \cite{xu_nips2013_1}.
AltProj and  RPCA-GD were implemented on the complete data matrix $\Y$. We must point out that we experimented with applying the batch methods for various sub-matrices of the data matrix, but the performance was not any better and thus we only report the results of the former method.
The other known parameters, $r$ for Alt-Proj, outlier-fraction for RPCA-GD, are set using the true values .
For both the techniques we set the tolerance as $10^{-6}$ and $100$ iterations (as opposed to the default $50$ iterations) to match that of ReProCS.

\subsubsection{Real Video Experiments}
We show comparisons on two real videos in this article. For extensive and quantitative comparisons done on the CDnet database, see \cite{rrpcp_review}. We show background recovery results for the Lobby and Meeting Room (or Curtain) dataset in Fig \ref{realvid}. 
Here we implement both online and batch algorithms in a similar manner and provide the comparison. Parameters set as $r=40$, $K=3$, $\alpha=20$, $\xi_t = \|\bpsi \lhat_{t-1}\|_2$ for ReProCS.